\documentclass [12pt, letterpaper]{article}
\pdfoutput=1
\usepackage{fullpage}
\usepackage{amsmath}
\usepackage{amssymb}
\usepackage{graphicx}
\usepackage{mathrsfs}
\usepackage{wrapfig}
\usepackage{boxedminipage}
\usepackage{epsfig}
\usepackage{setspace}
\usepackage{subfigure}
\usepackage{float}
\usepackage{hyperref}
\newcommand{\reef}[1]{(\ref{#1})}
\begin{document}

\begin{flushright}
\phantom{{\tt arXiv:0709.????}}
\end{flushright}

\bigskip
\bigskip
\bigskip

\begin{center} {\Large \bf Vortex and Droplet Engineering}
  
  \bigskip

{\Large\bf  in  }

\bigskip

{\Large\bf   Holographic Superconductors}

\end{center}

\bigskip \bigskip \bigskip \bigskip

\centerline{\bf Tameem Albash, Clifford V. Johnson}

\bigskip
\bigskip
\bigskip

  \centerline{\it Department of Physics and Astronomy }
\centerline{\it University of
Southern California}
\centerline{\it Los Angeles, CA 90089-0484, U.S.A.}

\bigskip

\centerline{\small \tt talbash,  johnson1,  [at] usc.edu}

\bigskip
\bigskip


\begin{abstract} 
\noindent 
We give a detailed account of the construction of non--trivial
localized solutions in a 2+1 dimensional model of superconductors
using a 3+1 dimensional gravitational dual theory of a black hole
coupled to a scalar field. The solutions are found in the presence of
a background magnetic field. We use numerical and analytic techniques
to solve the full Maxwell--scalar equations of motion in the
background geometry, finding condensate droplet solutions, and vortex
solutions possessing a conserved winding number. These solutions and
their properties, which we uncover, help shed light on key features of
the $(B,T$) phase diagram.
\end{abstract}
\newpage \baselineskip=18pt \setcounter{footnote}{0}


\section{Introduction}

An holographic model of some of the key phenomenological attributes of
superconductivity in 2+1 dimensions was proposed in
ref.\cite{Hartnoll:2008vx}. It works roughly as follows (a more
detailed review will follow in the next section).  The dual is a
simple model of gravity in four dimensions (with negative cosmological
constant) coupled to a $U(1)$ gauge field and a minimally coupled
charged complex scalar $\Psi$. Asymptotic values of the scalar on the
boundary correspond to the vacuum expectation value (vev) of a charged
operator in the 2+1 dimensional theory. For high temperatures
(relative to a scale set by non--zero charge density in the model) the
system is in a normal phase represented by a charged black hole
solution in the gravitational dual with the scalar set to zero. At a
critical temperature the system undergoes a phase transition, the
$U(1)$ getting spontaneously broken by a non--zero vev of the charged
operator\footnote{Strictly speaking, the $U(1)$ that is broken is
  global on the boundary, but it can be gauged in a number of ways
  without affecting the conclusions. See {\it e.g.}
  ref.\cite{Hartnoll:2008kx}.  Note that a global $U(1)$ does not restrict us to a spatially independent magnetic field.}. The gravitational description of this
is a charged black hole with a non--trivial scalar profile that gives
the vev on the boundary. This means that the black hole has ``scalar
hair'' in this regime. (For a discussion of violations of no--hair
theorems in this context, see ref.\cite{Gubser:2008px}. There, it was
shown that it is possible in this context, for large enough charges,
equivalent to the low temperature regime here.) The authors of
ref.\cite{Hartnoll:2008vx} showed using linear response theory that
the DC conductivity of this new phase diverges in a manner consistent
with the expectation that the system is in a superconducting
phase\footnote{Other holographic superconductors are available. See \emph{e.g.} refs.~\cite{Ammon:2008fc,Ammon:2009fe}.}. The authors carried out further study of the system in
ref.\cite{Hartnoll:2008kx}.

Our focus in this paper is the system in an external magnetic field,
continuing the work we began in ref.\cite{Albash:2008eh}.
Generically, for non--zero magnetic field $B$ filling the two spatial
dimensions, it is inconsistent to have non--trivial spatially
independent solutions on the boundary, and we present and study two
classes of localized solutions in some detail. The first is a
``droplet'' solution, the prototype of which was found in our earlier
work\cite{Albash:2008eh} as a strip in 2D (straightforwardly
generalized to circular symmetry in ref.\cite{Hartnoll:2008kx}), and
the second is a vortex solution, with integer winding number $\xi\in
{\mathbb{Z}}$, which is entirely new.  We obtain these as full solutions
of the Maxwell--scalar sector in a limit, and determine a number of
their properties.

Our analysis in various connected limits shows
where these solutions can exist in the $(B,T)$ plane. There is a
critical line below which droplets are not found, while vortices can
be found there. Our interpretation is that this region is the
superconducting phase, and that for non--zero $B$, the vortices
develop, trapping the magnetic flux into filaments, as is familiar in
type~II superconductors. Above the critical line, the system leaves
the superconducting phase, and either forms droplets of condensate or
simply reverts to the normal phase (dual to a dyonic black hole with
zero scalar everywhere, which may well yield lower action than the
droplets if we had back--reacting solutions to work with).

In section~2 we review the model, and discuss the two limits in which
most of our studies will be carried out. Section~3 reviews the
spatially independent solution corresponding to the prototype
superconducting solution. We briefly discuss our numerical approach to
finding the solution as a warmup for the more difficult problems in
the sequel. Section~4 presents our search for and construction of the
non--trivial spatially dependent solutions corresponding to condensate
droplets and to vortices. We discuss the numerical methods we used to
find them, and then examine a number of their properties. Section~5
examines aspects of the solutions' stability. We conclude in
section~6, and we also present two appendices. One appendix
establishes the normalisation of our gauge/gravity dual dictionary,
while the other discusses the flux quantization in our vortices.

\section{The Model, and Two Limits}
\subsection{The Model}
The holographic model of superconductivity in 2+1 dimensions proposed
in ref.\cite{Hartnoll:2008vx} is a  model of gravity in four
dimensions coupled to a $U(1)$ gauge field and a minimally coupled
charged complex scalar $\Psi$ with potential
$V(|\Psi|)=-2|\Psi|^2/L^2$. There is a negative cosmological constant
that defines a scale $L$ {\it via} $\Lambda=-3/L^2$.  The action is:
\begin{eqnarray} \label{eqt:EM_action}
S_{\mathrm{bulk}} = \frac{1}{2 \kappa_4^2}  \int d^4 x \sqrt{-G} \biggl\{ R + 
\frac{6}{L^2} +  L^2 \left(-\frac{1}{4} F^2 - \left| \partial \Psi - i g A \Psi \right|^2 - V \left(\left|\Psi \right| \right)\right) \biggr\}\ , 
\end{eqnarray}
where $\kappa_4^2=8\pi G_{\rm N}$ is the gravitational coupling and
our signature is $(-+++)$. 

We will use coordinates $(t, z, r,\phi)$
for much of our discussion, with $t$ time, $(r,\phi)$ forming a plane,
and $z$ a ``radial'' coordinate for our asymptotically AdS$_4$
spacetimes such that $z=0$ is the boundary at infinity. The AdS$_4$ metric is:
\begin{equation}
ds^2 = \frac{L^2}{z^2} \left( dz^2-dt^2 + dr^2 + r^2 d\phi^2 \right) \ .
\end{equation}
   Note that the
mass of the scalar $m^2_\Psi=-2/L^2$ is above the
Breitenlohner--Freedman stability bound\cite{Breitenlohner:1982bm}
$m^2_{\rm BF}=-9/4L^2$ for scalars in AdS$_4$.  We will write the scalar as:
\begin{equation}\label{realfields}
\Psi=\frac{{\tilde\rho}}{\sqrt{2}L}\exp(i\theta)\ .
\end{equation}
Near the boundary $z=0$ we have: 
\begin{equation}
{\tilde\rho}\to{\tilde\rho}_1 z+{\tilde\rho}_2 z^2\ , 
\end{equation}
where ${\tilde\rho}_i$ ($i=1,2$) sets the vacuum expectation value
(vev) of an operator ${\cal O}_i$ with dimension $\Delta=i$
\cite{Klebanov:1999tb}. Only one of these vevs can be non--zero at a
time, and we will choose to study the case of $i=1$, for much of the
paper. Our charged operator will be the order parameter for the
spontaneous breaking of the $U(1)$ symmetry. A gauge field of the form
$A=A_tdt$ does not give an electric field in the dual theory on
$(r,\phi)$, but defines instead\cite{Chamblin:1999tk} a $U(1)$ charge
density, $\rho$ and its conjugate chemical potential $\mu$, as we will
recall below.

Black holes in this study will be planar, {\it i.e.,} their horizons
are an $(r,\phi)$ plane at some finite $z=z_h$.  In the familiar
manner, their Hawking temperature $T$ and mass per unit horizon area
$\varepsilon=M/{\cal V}$, corresponds to the dual 2+1 dimensional
system at temperature $T$ and with energy
density~$\varepsilon$. Generically, the black hole will couple to the
gauge sector, having some profile for the field $A_t$. The temperature
$T$ will have dependence on the charge density parameter~$\rho$. This
is quite natural since without the charge density there is no other
scale in the theory, and there would be no meaning to a high or low
temperature phase, and hence no possibility of a phase transition.

The high $T$ phase of the theory is simply the charged black hole
(Reissner--Nordstr\"{o}m (AdS--RN)) with the scalar $\Psi$
vanishing. This corresponds to the non--superconducting or ``normal''
phase of the theory, where the order parameter vanishes.  The mass of
the scalar $\Psi$ is set not just by $V(|\Psi|)$ but by the density
$\rho$ through the coupling to the gauge field. In fact $m^2_\Psi$
decreases with $T$ until at~$T_c$ it goes below $m^2_{\rm BF}$,
becoming tachyonic. The theory seeks a new solution, in which the
black hole is no longer AdS--RN, but one that has a non--trivial
profile for~$\Psi$.

In studying the system in a magnetic field background, there are some
generic expectations to consider.  The magnetic field $B$ (which fills
the two dimensions of the superconducting theory), also contributes to
$m^2_{\Psi}$, {\it via} its square, but contributes with {\it opposite
  sign} to the electric contribution of the background.  It therefore
lowers the temperature $T_c$ at which $m^2_\Psi$ falls below $m^2_{\rm
  BF}$, triggering the phase transition.  On these grounds alone one
then (naively) expects a critical line in the $(B,T)$ plane connecting $(0,T_c)$ to some $(B_c,0)$, but it is important to determine exactly what physics lies on either side of the
line. Our solutions and our study of their properties help in
establishing some of this. The solutions can be found by
solving equations of motion in certain limits and we remind the reader
of them in the next two subsections.

\subsection{The Decoupling Limit}
\label{sec:decoupling}
After a field redefinition:
\begin{equation}
A_\mu \to \frac{1}{g} A_\mu \ , \quad \Psi \to \frac{1}{g} \Psi \ ,
\end{equation}
our action~(\ref{eqt:EM_action}) becomes:
\begin{equation}
S_{\mathrm{bulk}} = \frac{1}{2 \kappa_4^2} \int d^4 x \sqrt{-G} \left\{ R + \frac{6}{L^2} + \frac{L^2}{g^2} \left(-\frac{1}{4} F^2 - \left| \partial \Psi - i A \Psi \right|^2 + \frac{2}{L^2} \bar{\Psi} \Psi \right) \right\}  \ .
\end{equation}
If we consider the limit $g \to \infty$, then the Maxwell--scalar
sector decouples from gravity.  This allows us to work with a fixed
uncharged background, which we take to be the AdS$_4$--Schwarzschild
(AdS--Sch) black hole, given by:
\begin{equation}
ds^2 = \frac{L^2 \alpha^2}{z^2} \left( - f \left(z \right) dt^2 + dr^2 + r^2 d\phi^2 \right) + \frac{L^2}{z^2} \frac{1}{f \left(z \right)} dz^2 \ ,
\end{equation}
where $f \left(z \right) = 1- z^3$.  The coordinate $z$ is a
dimensionless parameter scaled such that the event horizon is at $z_h
= 1$.  The Hawking temperature is given by the usual Gibbons--Hawking
calculus \cite{Gibbons:1979xm}:
\begin{equation}
T = \frac{3}{4 \pi} \alpha \ .
\end{equation}
Note that $\alpha$ is related to the mass of the black hole:
\begin{equation}
\varepsilon=\frac{M}{{\cal V}}=\frac{L^2 \alpha^3}{\kappa_4^2}\ ,
\end{equation}
where ${\cal V}$ is the volume of the $(r,\phi)$ plane. In terms of
the two real fields $({\tilde\rho},\theta)$ into which we decomposed
$\Psi$ into in equation~(\ref{realfields}) we have:
\begin{equation} \label{eqt:action_rho}
- L^2\left| \partial \Psi - i A \Psi \right|^2 +2 \bar{\Psi} \Psi  = -\frac{1}{2} G^{\mu \nu} \left[ \partial_\mu {\tilde\rho} \partial_\nu {\tilde\rho} + {\tilde\rho}^2 \left( \partial_\mu \theta \partial_\nu \theta - 2 A_\mu \partial_\nu \theta + A_\mu A_\nu \right)\right] + \frac{1}{L^2} {\tilde\rho}^2 \ .
\end{equation}
From the action, we derive the equation of motion for the fields
${\tilde\rho}$, $\theta$, and $A_\mu$:
\begin{eqnarray} \label{eqt:eom}
& \frac{1}{\sqrt{-G}} \partial_\mu \left( \sqrt{-G} G^{\mu \nu} \partial_\nu {\tilde\rho} \right) - G^{\mu \nu} {\tilde\rho} \left( A_\mu - \partial_\mu \theta \right) \left( A_\nu - \partial_\nu \theta \right) + \frac{2}{L^2} {\tilde\rho} = 0 \ , & \nonumber \\
& - \frac{1}{\sqrt{-G}} \partial_\mu \left(\sqrt{-G} G^{\mu \nu} {\tilde\rho}^2 \left(A_\nu - \partial_\nu \theta \right)\right) = 0 \ , & \\
& \frac{1}{\sqrt{-G}} \partial_\nu \left(\sqrt{-G} G^{\nu \lambda } G^{\mu \sigma} F_{\lambda \sigma} \right)  - \frac{G^{\mu \nu}}{L^2} {\tilde\rho}^2 \left( A_\nu - \partial_\nu \theta \right)= 0 \ . & \nonumber
\end{eqnarray}
While the background itself will have no charge in this limit, there
will of course still be a non--trivial gauge field $A$, and for an
electric background $A=A_t dt$, we will have, as $z\to 0$:
\begin{equation}
 \frac{ A_t}{\alpha}\equiv \tilde{A}_t \to \mu - \rho \ z \ ,
\end{equation}
defining a chemical potential $\mu$ and a charge density $\rho$.

\subsection{The Probe Limit}
\label{sec:probe}
Sometimes we will also work in a probe limit, where we take the scalar
in the Maxwell--scalar sector to be small, and hence not
back--reacting on either the geometry. In general, we can do this at
arbitrary $g$. (We will combine this with the decoupling limit
($g\to\infty$) for one case, as we shall see later.) For finite $g$ we will consider our small non--backreacting scalar to
be moving in a dyonic Reissner--Nordstr\"{o}m background, given by
\cite{Romans:1991nq}:
\begin{eqnarray} \label{eqt:dyonic_metric}
ds^2 &=& \frac{L^2 \alpha^2}{z^2} \left( - f \left(z \right) dt^2 + dr^2 + r^2 d \phi^2 \right) +\frac{L^2}{z^2} \frac{d z^2}{f \left(z \right)} \ , \\
F &=& 2 h \alpha^2 r dr \wedge d \phi + 2 q \alpha dz \wedge dt\ , \nonumber \\
f \left(z \right) &=& 1 + \left( h^2 + q^2 \right) z^4 - \left(1 + h^2 + q^2 \right) z^3 = \left(1-z \right)\left( z^2 +z+ 1 - \left(h^2 +q^2\right) z^3 \right) \ . \nonumber 
\end{eqnarray}
The temperature  and charge density are given by:
\begin{equation} \label{eqt:temp}
T = \frac{1}{\beta} =  \frac{\alpha}{4 \pi} \left( 3 - h^2 - q^2 \right) \ ,\qquad \rho =\frac{1}{\mathcal{V} \beta} \frac{\delta S_{\mathrm{on-shell}}}{\delta A_t (z = 0)} = - \frac{L^2}{\kappa_4^2} q \alpha^2 \ .
\end{equation}
We choose a gauge such that the gauge field is written as:
\begin{equation}
A = h \alpha^2 r^2 d \phi + 2 q \alpha \left( z - 1 \right) d t \ .
\end{equation}
%

%
\section{Spatially Independent Solution} \label{sec:spatially_independent}
%
We begin by considering a spatially independent solution, reviewing
the original presentation of ref.\cite{Hartnoll:2008vx}, working in
the decoupling limit of section~\ref{sec:decoupling}. We take an
ansatz for the fields given by:
\begin{equation}
\theta \equiv \mathrm{const} \ , \quad  {\tilde\rho}=\tilde{\rho} \left( z \right) \ , \quad A_t \equiv \alpha \tilde{A}_t \left( z \right) \ , \quad A_\phi = 0 \ .
\end{equation}
where $\tilde{\rho}$ and $\tilde{A}_t$ are dimensionless fields.  The equations of motion are given by:
\begin{eqnarray} \label{eqt:eom1}
& \partial_z^2 \tilde{\rho}  + \left(\frac{f'}{f}-\frac{2}{z} \right) \partial_z \tilde{\rho} + \frac{1}{  f^2  } \tilde{\rho} \tilde{A}_t^2 + \frac{2}{z^2 f}\tilde{\rho}= 0 \ , & \nonumber \\
&\partial_z^2 \tilde{A}_t - \frac{1}{z^2 f} \tilde{\rho}^2 \tilde{A}_t = 0 \ . & 
\end{eqnarray}
We can study the equations' behaviour near the event horizon ({\it i.e.} as $z \to 1$):
\begin{eqnarray}
& \left[ \partial_z \tilde{\rho} - \frac{2}{3} \tilde{\rho} \right]_{z=1} = 0 \ , \quad \left[ \partial_z^2 \tilde{\rho} - \frac{5}{6} \partial_z \tilde{\rho} + \tilde{\rho} + \frac{1}{18} \tilde{\rho} \left( \partial_z \tilde{A}_t \right)^2 \right]_{z=1}  = 0 \ , & \nonumber \\
& \tilde{A}_t \Big|_{z=1} = 0 \ , \quad \left[ \partial_z^2 \tilde{A}_t + \frac{1}{3} \tilde{\rho}^2 \partial_z \tilde{A}_t \right]_{z=1} = 0  \ . &\nonumber
\end{eqnarray}
We note that the only free variables (to be chosen) at the event horizon are $\tilde{\rho} (1)$ (or $\partial_z \tilde{\rho} (1)$) and $\partial_z \tilde{A}_t (1)$.  The other limit to study is to consider the behavior near the AdS boundary ($z \to 0$):
\begin{eqnarray}
 \left[ \partial_z^2 \tilde{\rho} - \frac{2}{z} \partial_z \tilde{\rho} + \frac{2}{z^2} \tilde{\rho} \right]_{z=0} = 0 \ , \qquad \left[ \partial_z^2 \tilde{A}_t - \frac{1}{z^2} \tilde{\rho}^2 \tilde{A}_t \right]_{z=0} = 0 \ ,  \nonumber
\end{eqnarray}
which has as solutions
\begin{equation} \label{eqt:rho_BC}
\tilde{\rho} ( z \to 0) \to \tilde{\rho}_1  z + \tilde{\rho}_2 z^2  \ , \quad \tilde{A}_t ( z \to 0) \to \mu - \rho \ z \ ,
\end{equation}
where $\tilde{\rho}_1$, $\tilde{\rho}_2$, $\mu$, and $\rho$ are
constants related to the vev of a $\Delta = 1$ operator, the vev of a
$\Delta = 2$ operator, the chemical potential, and the charge density
of the dual field theory respectively.  The solution for
$\tilde{\rho}$ at the AdS boundary admits two normalisable modes, and
therefore the constants are associated with vevs of two separate
operators.  Only one of these vevs is to be non--zero at a time, and
the two different gauge theories are related to each other \emph{via}
a Legendre transformation \cite{Klebanov:1999tb}.
\subsection{Numerical Analysis}
\begin{figure}[h] 
   \centering
   \includegraphics[width=2.5in]{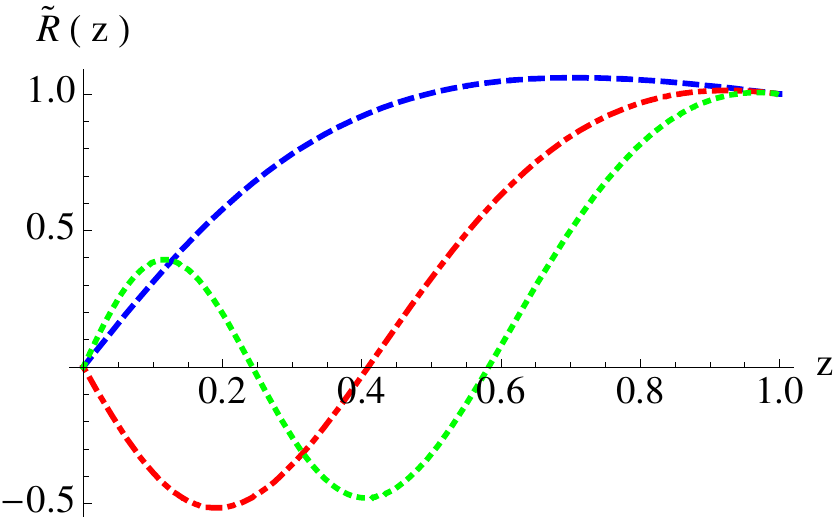} 
   \caption{\small Three solutions with the same $\tilde{R}(1)$ but different $\partial_z \tilde{A}_t (1)$ that satisfy Dirichlet boundary conditions at $z=0$.  The solutions are distinguished by the number of nodes (times they cross the $z$--axis) they have.}
   \label{fig:constant_z_behavior}
\end{figure}

\begin{figure}[ht]
\begin{center}
\subfigure[$$]{\includegraphics[width=2.5in]{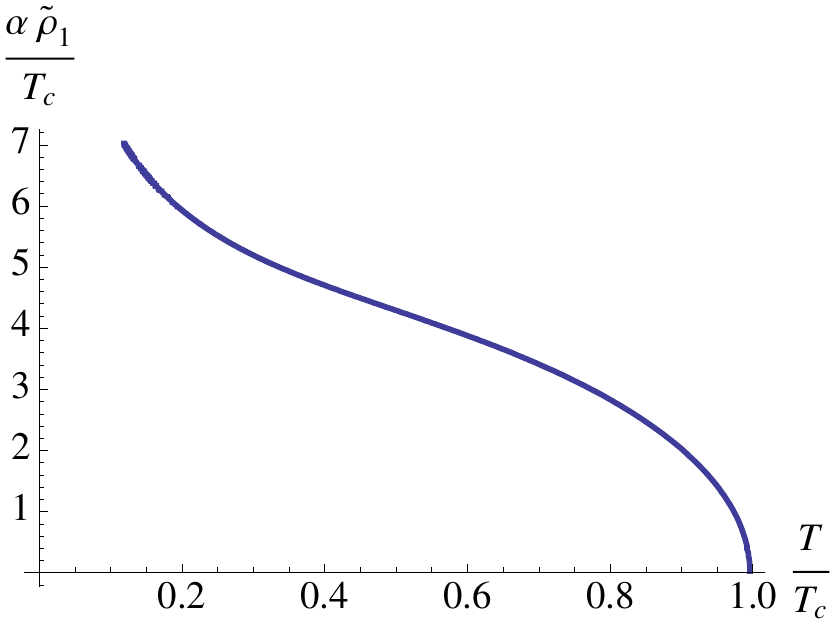}\label{fig:constant_sol_rho1}} \hspace{0.5cm}
\subfigure[$$]{\includegraphics[width=2.5in]{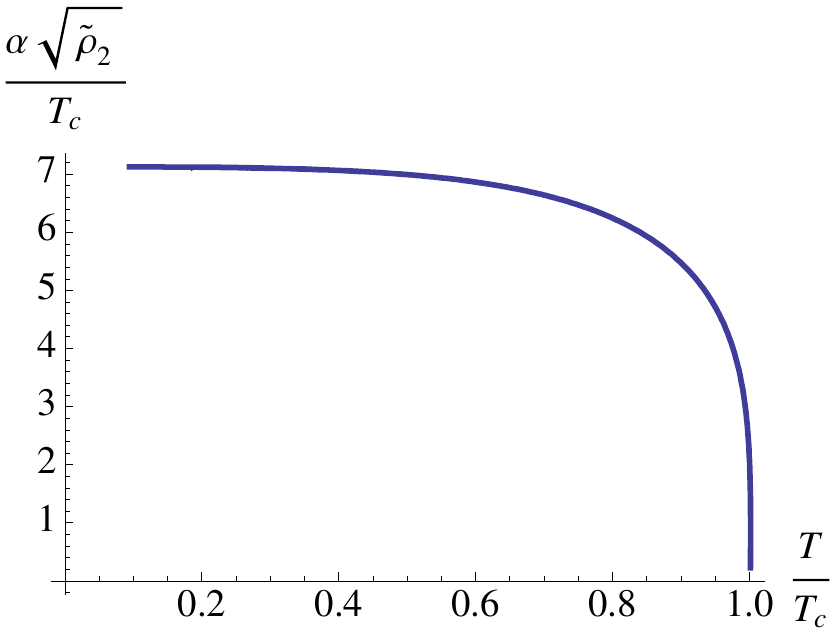}\label{fig:constant_sol_rho2}}\hspace{0.5cm}
\caption{\small  Vaccuum expectation values for the scalar.  Here $T_c$ is defined to be $0.226 \alpha \sqrt{\rho}$ and $0.118\alpha  \sqrt{\rho}$ for the $\Delta = 1$ and $\Delta = 2$ operator respectively.} \label{fig:constant_solution}
\end{center}
\end{figure}

To simplify the numerical analysis, it is convenient to define a new
field $\tilde{R}(z)$ such that:
\begin{equation}
\tilde{R}(z) = z \tilde{\rho}(z)\ .
\end{equation}
With this redefinition, the boundary condition of having either
$\tilde{\rho}_1$ or $\tilde{\rho}_2$ in equation \reef{eqt:rho_BC} to
be zero becomes the requirement of having either a Dirichlet or a
Neumann boundary condition on $\tilde{R}$ at the AdS boundary.  The
equations in the bulk of AdS are given by:
\begin{eqnarray}
& \partial_z^2 \tilde{R} + \frac{f'}{f} \partial_z \tilde{R} + \frac{1}{f^2} \tilde{R} \tilde{A}_t^2 + \left( \frac{f'}{z f} - \frac{2}{z^2} + \frac{2}{z^2 f} \right) \tilde{R} = 0 \ , & \nonumber  \\
& \partial_z^2 \tilde{A}_t - \frac{1}{f} \tilde{R}^2 \tilde{A}_t = 0 \ , & 
\end{eqnarray}
and we solve them using a shooting method (discretizing using finite
differences) with shooting conditions:
\begin{eqnarray}\label{eq:conditions}
& \tilde{R}(1) = \mathrm{const} \ , \quad \partial_z \tilde{R}(1) = -\frac{1}{3} \tilde{R}(1) \ , \quad \partial_z^2 \tilde{R} (1) = - \frac{5}{18} \tilde{R}(1) - \frac{1}{18} \left(\partial_z \tilde{A}_t(1)  \right)^2 \tilde{R}(1) \ . & \nonumber \\
& \tilde{A}_t(1) = 0 \ , \quad \partial_z \tilde{A}_t (1) = \mathrm{const} \ , \quad \partial_z^2 \tilde{A}_t (1) = - \frac{1}{3} \tilde{R}(1)^2 \partial_z \tilde{A}_t (1) \ . &
\end{eqnarray}
The solution at $z=0$ goes as:
\begin{equation}
\tilde{R} (z \to 0) = \tilde{R}_1 + \tilde{R}_2 \ z \ , \quad \tilde{A}_t ( z \to 0 )  = \mu - \rho \ z \ .
\end{equation}
We fix $\tilde{R}(1)$ and then tune $\partial_z \tilde{A}_t (1)$ until
the solution satisfies the necessary Dirichlet or Neumann boundary
condition at $z = 0$. We then read off the scalar and also the value
of~$\rho$ for that solution, which defines the temperature. We can
determine $T_c$ since there is a minimum charge density (over
temperature squared) needed for the scalar field to condense. Note
that there are multiple choices for $\partial_z \tilde{A}_t(1)$ that
give the necessary boundary condition at the AdS boundary, sample
solutions of which we present in figure \ref{fig:constant_z_behavior}.
Solutions with a greater number of nodes are associated with higher
chemical potential/charge density.  These solutions are of a higher
energy and so are thermodynamically unfavorable, therefore we only
present results of the zero--node solutions in what follows.
In figure \ref{fig:constant_solution} we show the solutions for the
scalar values ${\tilde \rho}_{1,2}$ at the boundary which give the
vevs of the operators ${\cal O}_{1,2}$.  As anticipated, in each case,
the vev of the operator is zero above $T/T_c = 1$.  Below $T/T_c = 1$, it is not zero, showing the spontaneous breaking of the $U(1)$ symmetry.
%
\section{Spatially Dependent Solutions}
%
A non--zero magnetic field $B$ in the $(r,\phi)$ plane will correspond
to some non--zero $A_\phi(r)$. In such a case, consistency of the
solution requires the fields to have some spatial dependence in the
plane.  This situation was studied in the linear case in
ref.~\cite{Albash:2008eh}, (see also ref.\cite{Hartnoll:2008kx}) but
here we consider the full non--linear problem of equation
\reef{eqt:eom}.  First, notice that the $U(1)$ gauge transformation
acts as:
\begin{equation}\label{eq:gauge}
\rho \to \rho \ , \quad \theta  \to \theta + \Lambda \ , \quad A_\mu  \to  A_\mu  + \partial_\mu \Lambda\ . 
\end{equation}
In the previous section we chose $\theta$ to have no non--trivial
dependence. Naively, it would seem that we can freely shift $\theta$
by gauge transformations.  However, this freedom is only available if
the gauge symmetry is not broken. We will return to this once we have
constructed the solutions.  This motivates us to consider the
following ansatz:
\begin{equation} \label{eqt:ansatz}
\theta \equiv \zeta + \xi \ \phi \ , \quad {\tilde \rho} = \tilde{\rho} \left( \tilde{r} , z \right) \ , \quad A_t = \alpha \tilde{A}_t \left( \tilde{r} , z \right) \ , \quad A_\phi = \tilde{A}_\phi \left(\tilde{r} , z \right) \ ,
\end{equation}
where we have defined a dimensionless radial coordinate $\tilde{r} =
\alpha r$, dimensionless fields $\tilde{\rho}$, $\tilde{A}_t$, and
$\tilde{A}_\phi$, and $\left( \zeta, \xi \right)$ are constants where
$\xi$ is an integer.  Under this ansatz, the equations of motion
reduce to:
\begin{eqnarray} \label{eqt:eom2}
& \partial_z^2 \tilde{\rho}  + \left(\frac{f'}{f}-\frac{2}{z} \right) \partial_z \tilde{\rho} + \frac{1}{f  } \left( \partial_{\tilde{r}}^2 \tilde{\rho} + \frac{1}{\tilde{r}} \partial_{\tilde{r}} \tilde{\rho} - \frac{1}{\tilde{r}^2} \tilde{\rho} \left(\tilde{A}_\phi-\xi \right)^2 \right) + \frac{1}{  f^2  } \tilde{\rho} \tilde{A}_t^2 + \frac{2}{z^2 f}\tilde{\rho}= 0 \ , & \nonumber \\
& \partial_z^2  \tilde{A}_\phi  + \frac{f'}{f} \partial_z \tilde{A}_\phi+\frac{1}{f }   \left(\partial_r^2 \tilde{A}_\phi - \frac{1}{\tilde{r}} \partial_{\tilde{r}} \tilde{A}_\phi \right) - \frac{1}{z^2 f }  \tilde{\rho}^2 \left(\tilde{A}_\phi -\xi \right)= 0 \ , & \\
& \partial_z^2 \tilde{A}_t + \frac{1}{ f} \left( \partial_{\tilde{r}}^2 \tilde{A}_t + \frac{1}{\tilde{r}} \partial_{\tilde{r}} \tilde{A}_t \right) - \frac{1}{z^2 f} \tilde{\rho}^2 \tilde{A}_t = 0 \  ,& \nonumber
\end{eqnarray}
where the equation of motion for the field $\theta$ is trivially
satisfied by our ansatz.  Near the event horizon, these equations
reduce to the following conditions that must be satisfied:
\begin{eqnarray} \label{eqt:eom_horizon}
& \left[\partial_{\tilde{r}}^2 \tilde{\rho}+ \frac{1}{\tilde{r}} \partial_{\tilde{r}} \tilde{\rho} - \frac{1}{\tilde{r}^2} \tilde{\rho} \left(\tilde{A}_\phi - \xi \right)^2 +2 \tilde{\rho} =  3 \partial_z \tilde{\rho} \right]_{z=1} \  , \quad  \left[ \partial_z^2 \tilde{\rho} = - \frac{4}{3} \tilde{\rho} -  \frac{1}{9} \tilde{\rho} \left( \partial_z \tilde{A}_t \right)^2 \right]_{z=1}  \ , & \nonumber \\
&\left[ \partial_{\tilde{r}}^2 \tilde{A}_\phi - \frac{1}{\tilde{r}} \partial_{\tilde{r}} \tilde{A}_\phi - \left(\tilde{A}_\phi- \xi \right) \tilde{\rho}^2  =  3 \partial_z \tilde{A}_\phi \right]_{z=1}  \ , \quad \left[ \partial_z^2 \tilde{A}_\phi =  \frac{2}{3} \left(\tilde{A}_\phi -\xi \right) \tilde{\rho}^2 - 2 \partial_z \tilde{A}_\phi \right]_{z=1} \  , & \nonumber\\
& \tilde{A}_t (\tilde{r}, z=1) = 0 \ , \quad \left[ 3 \partial_z^2 \tilde{A}_t = \partial_z \left( \partial_{\tilde{r}}^2 \tilde{A}_t + \frac{1}{\tilde{r}} \partial_{\tilde{r}} \tilde{A}_t - \tilde{\rho}^2 \tilde{A}_t \right) + 2 \tilde{A}_t \tilde{\rho}^2 \right]_{z=1}  \ , & 
\end{eqnarray}
where in the first equation we have used that $\tilde{A}_t (\tilde{r},
z=1) = 0$. We now have three free functions to fix in these equations,
$\partial_z \tilde{\rho} (\tilde{r}, z=1), \partial_z \tilde{A}_\phi
(\tilde{r}, z=1),$ and $ \partial_z \tilde{A}_t (\tilde{r}, z=1)$,
which determine the spatial profile of the solutions at the event
horizon.
Note that in order to avoid a divergence in the equation for
$\tilde{\rho}$ at $\tilde{r}=0$, we must have that near $\tilde{r}=0$,
the field~$\tilde{\rho}$ must go as $\tilde{r}^\xi$.  This motivates
the following field redefinitions:
\begin{equation}
\tilde{\rho} = z \tilde{r}^{\xi} \tilde{R} (\tilde{r}, z) \ , \quad \tilde{A}_\phi = \tilde{r}^2 \tilde{A} \ ,
\end{equation}
where $\tilde{R}$ near $\tilde{r} = 0$ is a non--zero value.  The
particular redefinition of $\tilde{A}_\phi$  simplifies the
numerical analysis.  Our  new equations of
motion for these fields are:
\begin{eqnarray} \label{eqt:eom_xi}
 \partial_z^2 \tilde{R}  + \frac{f'}{f} \partial_z \tilde{R} + \frac{1}{f  } \left( \partial_{\tilde{r}}^2 \tilde{R} + \frac{2\xi + 1}{\tilde{r}} \partial_{\tilde{r}} \tilde{R} + \frac{\xi^2}{\tilde{r}^2} \tilde{R} - \frac{1}{\tilde{r}^2} \tilde{R} \left(\tilde{r}^2 \tilde{A}-\xi \right)^2 \right)&& \nonumber\\
+ \frac{1}{  f^2  } \tilde{R} \tilde{A}_t^2 + \left(\frac{f'}{z f} - \frac{2}{z^2} + \frac{2}{z^2 f}\right) \tilde{R}&=& 0 \ ,  \nonumber \\
 \partial_z^2  \tilde{A}  + \frac{f'}{f} \partial_z \tilde{A}+\frac{1}{f }   \left(\partial_r^2 \tilde{A} + \frac{3}{\tilde{r}} \partial_{\tilde{r}} \tilde{A} \right) - \frac{\tilde{r}^{2 \xi}}{ f }  \tilde{R}^2 \left(\tilde{A} - \frac{\xi}{\tilde{r}^2} \right)&=& 0 \ ,  \nonumber \\
 \partial_z^2 \tilde{A}_t + \frac{1}{ f} \left( \partial_{\tilde{r}}^2 \tilde{A}_t + \frac{1}{\tilde{r}} \partial_{\tilde{r}} \tilde{A}_t \right) - \frac{\tilde{r}^{2 \xi}}{ f} \tilde{R}^2 \tilde{A}_t &=& 0 \  .
\end{eqnarray}
\begin{eqnarray} \label{eqt:eom_horizon_xi}
&& \left[\partial_{\tilde{r}}^2 \tilde{R}+ \frac{2 \xi + 1}{\tilde{r}} \partial_{\tilde{r}} \tilde{R}  + \frac{\xi^2}{\tilde{r}^2} \tilde{R} - \frac{1}{\tilde{r}^2} \tilde{R} \left(\tilde{r}^2 \tilde{A} - \xi \right)^2 - \tilde{R} =  3 \partial_z \tilde{R} \right]_{z=1} \  ,  \nonumber\\
&& \hskip1cm \left[ \partial_z^2 \tilde{R} = - \frac{1}{3} \tilde{R} -  2 \partial_z \tilde{R} - \frac{1}{9} \tilde{R} \left( \partial_z \tilde{A}_t \right)^2 \right]_{z=1}  \ ,  \nonumber \\
&&\left[ \partial_{\tilde{r}}^2 \tilde{A} + \frac{3}{\tilde{r}} \partial_{\tilde{r}} \tilde{A} - \left(\tilde{A}- \frac{\xi}{\tilde{r}^2} \right) \tilde{r}^{2 \xi} \tilde{R}^2  =  3 \partial_z \tilde{A}_\phi \right]_{z=1}  \ , \quad \left[ \partial_z^2 \tilde{A} =  - 2 \partial_z \tilde{A} \right]_{z=1} \  ,  \\
&&\left[ \partial^2_{\tilde{r}} \tilde{T} + \frac{1}{\tilde{r}}\partial_{\tilde{r}} \tilde{T} - \tilde{r}^{2 \xi} \tilde{R}^2 \tilde{T} = 6 \partial_z \tilde{T} \right]_{z=1} \ , \quad  \left[ \partial_z^2 \tilde{T} =  - 2 \partial_z \tilde{T} \right]_{z=1} \ ,  \nonumber 
\end{eqnarray}
where we have used that:
\begin{equation}
\lim_{z \to 1} A_t (\tilde{r},z) = \lim_{z \to 1}  \left(1- z \right) \tilde{T} (\tilde{r},z)\ .
\end{equation}
In particular, at the event horizon, we can expand the fields near $\tilde{r}=0$ as:
\begin{eqnarray} \label{eqt:expansion}
\lim_{\tilde{r} \to 0} \tilde{R} \left(\tilde{r}, z \right) & = & R_0(z) \left( 1 + \frac{1}{2} a_2 \tilde{r}^2  + O(\tilde{r}^3) \right) \ , \nonumber \\
\lim_{\tilde{r} \to 0} \tilde{A}  \left(\tilde{r}, z \right) & = & A_0(z) \left( 1 + \frac{1}{2} b_2 \tilde{r}^2 + O(\tilde{r}^3) \right)  \ , \\
\lim_{\tilde{r} \to 0} \tilde{T} \left(\tilde{r}, z \right) & = & T_0(z) \left(1+  \frac{1}{2} c_2 \tilde{r}^2 + O(\tilde{r}^3) \right) \ .\nonumber 
\end{eqnarray}
Substitution into the equation of motion at the horizon yields:
\begin{eqnarray}
 a_2 &=& \frac{ R_0(1) + 3 \partial_z R_0 (1)  - 2 \xi A_0(1) R_0(1)}{2(\xi + 1) R_0(1)} \ , \nonumber \\
 b_2 &=& \left\{
 \begin{array}{cc} \frac{A_0(1) R_0(1)^2 + 3 \partial_z A_0(1)}{4 A_0(1)}  \ , & \xi  = 0\\
 \frac{ - \xi R_0(1)^2 + 3 \partial_z A_0(1)}{4 A_0(1)}  \ , & \xi = 1 \\
 \frac{ 3 \partial_z A_0(1)}{4 A_0(1)} \ , & \xi \ge 2
 \end{array} \right. \ ,
 \\
 c_2 &=& \left\{ \begin{array} {cc}
 \frac{T_0(1) R_0(1)^2 + 6 \partial_z T_0(1)}{2 T_0(1)} \ , &\ \xi = 0 \\
  \frac{6 \partial_z T_0(1)}{2 T_0(1)} \ , &\ \xi \ge 1
 \end{array} \right.
 \ . \nonumber
\end{eqnarray}
The solutions we study are characterized by the value of $\xi$ and the
$\tilde{r}$ asymptotic behavior of the field $\tilde{R}$.  For any
allowed value of $\xi$, the solution for $\tilde{R}$ can asymptote to
zero or to a constant non--zero value.  

We will  be extracting non--trivial profiles for the fields at the
boundary at $z=0$ as follows:
\begin{eqnarray}
 \tilde{A}_t(\tilde{r},z) &=& \mu(\tilde{r}) - \rho(\tilde{r}) z \ , \nonumber \\
\tilde{A}_\phi(\tilde{r},z) \equiv \tilde{r}^2 \tilde{A}(\tilde{r},z) &=& a_\phi(\tilde{r}) + J_\phi(\tilde{r}) z \ , \nonumber \\
\tilde{\rho}(\tilde{r},z) \equiv z \tilde{R} (\tilde{r},z)&=& \tilde{R}(\tilde{r},0)z + \partial_z \tilde{R}(\tilde{r},0) z^2 \ ,
\end{eqnarray}
where $\mu$ is related to the chemical potential, $\rho$ is related to
the charge density, $J_\phi$ is related to the azimuthal current, and
$a_\phi$ is related to the magnetic field {\it via} $\tilde{B}_z
=( \partial_{\tilde r} a_\phi ) / {\tilde r}$.  For the exact
relationships, please consult Appendix \ref{sec:Dictionary}.

\subsection{Droplet Solutions}
We consider the case of the case of $\xi=0$, and use ${\cal O}_1$ as our order parameter.  For this choice of $\xi$ the solution that asymptotes to a constant value is simply the
spatially--independent solution described earlier in
section~\ref{sec:spatially_independent}. In this section we consider
solutions that asymptote to zero.  To that end, we fix the following
functions to:
\begin{equation}
\partial_z \tilde{R} (\tilde{r}, z=1) = - \frac{1}{3} \left(1 + \gamma \right) \tilde{R}(\tilde{r},1) \ , \quad \partial_z \tilde{A}(\tilde{r},1) = 0 \ ,
\end{equation}
where $\gamma$ is a positive number.  With these choices, the equations of motion at the event horizon reduce to:
\begin{eqnarray} \label{eqt:eom_droplet_horizon}
& \left[\partial_{\tilde{r}}^2 \tilde{R} + \frac{1}{\tilde{r}} \partial_{\tilde{r}} \tilde{R} -\tilde{r}^2 \tilde{A}^2 \tilde{R} + \gamma \tilde{R} \right]_{z=1}  = 0  \ , \quad \left[ \partial_z^2 \tilde{R} = \left(\frac{1}{3} + \frac{2}{3}  \gamma \right) \tilde{R} - \frac{1}{9} \tilde{R} \tilde{T}^2 \right]_{z=1} \ , & \nonumber \\
& \left[ \partial_{\tilde{r}}^2\tilde{A} + \frac{3}{\tilde{r}} \partial_{\tilde{r}} \tilde{A}  -   \tilde{R}^2 \tilde{A}  \right]_{z=1} = 0 \ , \quad \left[ \partial_z^2 \tilde{A}  \right]_{z=1} = 0 \ ,& \\
&  \left[ \partial_{\tilde{r}}^2 \tilde{T} + \frac{1}{\tilde{r}} \partial_{\tilde{r}} \tilde{T}  -\tilde{R}^2   \tilde{T}  =  6 \partial_z \tilde{T} \right]_{z=1} \ , \quad \left[ \partial_z^2 \tilde{T} = -  2 \partial_z \tilde{T} \right]_{z=1}  \ . &  \nonumber
\end{eqnarray}
The coefficients in equation \reef{eqt:expansion} are given by:
\begin{equation} 
a_2 = -\frac{\gamma}{2} \ , \quad  b_2 = \frac{R_0(1)^2}{4} \ , \quad c_2 = \frac{T_0(1) R_0(1)^2 +6 \partial_z T_0(1)}{2 T_0(1)} \ .
\end{equation}
%

\subsubsection{Numerical Procedures} \label{sec:NP}
%
We begin by solving equations \reef{eqt:eom_horizon_xi}, which are at
the event horizon.  For a given $R_0(1)$, we find that there is a
specific value for $A_0(1)$ and $T_0(1)$ that gives regular solutions
for the three functions $\tilde{R}(\tilde{r},1), \
\tilde{A}(\tilde{r},1), \ \tilde{T}(\tilde{r},1)$.  The coupled
ordinary differential equations are discretized using an explicit
finite difference method, and we determine the values of $A_0(1)$ and
$T_0(1)$ using a shooting method.  By this we mean that we pick values
for $A_0(1)$ and $T_0(1)$ at the origin and ``shoot'' towards
$\tilde{r} \to \tilde{r}_{\mathrm{max}}$, where ${\tilde
  r}_{\mathrm{max}}$ is the largest radius out to which we will
construct our solutions.  Typically, this leads to a divergence in the
functions, and therefore we iterate the procedure, fine--tuning our
initial conditions such that a regular solution is found. This has now
determined our initial conditions for the bulk problem.

The initial conditions at the event horizon having been determined, we
solve the bulk equations of motion \reef{eqt:eom_xi} and shoot towards
the boundary at $z=0$.  The coupled partial differential equations are
discretized using a finite difference method, and we adjust the mesh
spacings $\Delta z$ and $\Delta r$ until we achieve stability for our
code.

In order to satisfy the necessary boundary conditions at the AdS
boundary, we try to minimize the {\it positive} area under the curve
of $\partial_z \tilde{R} (\tilde{r},0)$.  We accomplish this
minimization by fine tuning our choice of $\partial_z T (\tilde{r},1)$
at the event horizon. This is acheived by expanding it in an
appropriate basis of functions in ${\tilde r}$ and using a
Monte--Carlo method to determine the coefficients (with the area
playing the role of energy).

\subsubsection{Sample Solutions}

\begin{figure}[h!]
\begin{center}
\subfigure[Scalar Field for $\gamma =0.5$]{\includegraphics[width=2.25in]{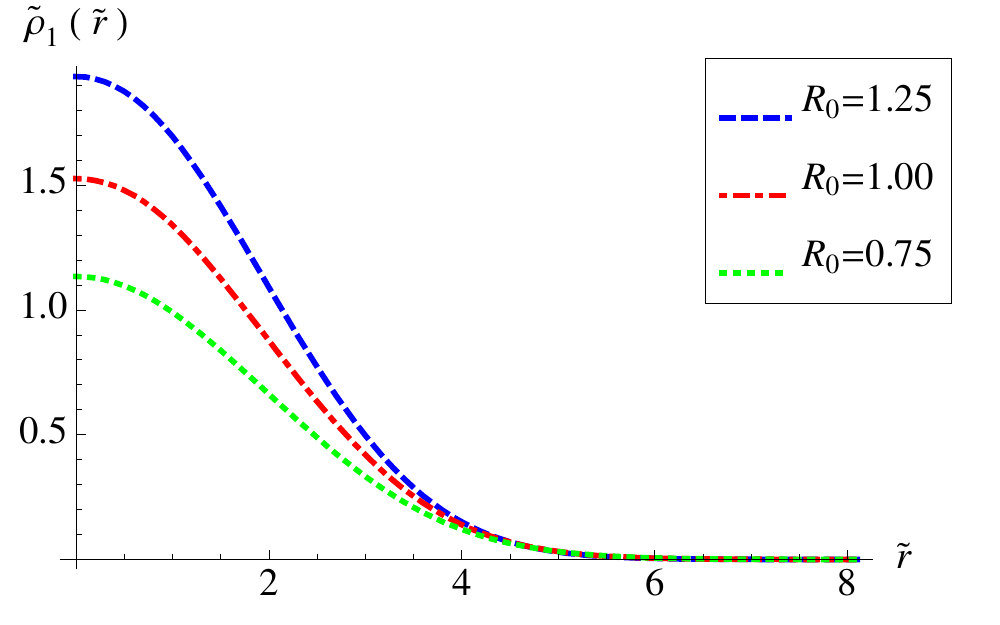}\label{fig:droplet_rho_gamma=0.5}} \hspace{0.5cm}
\subfigure[Scalar Field for $\gamma =2$]{\includegraphics[width=2.25in]{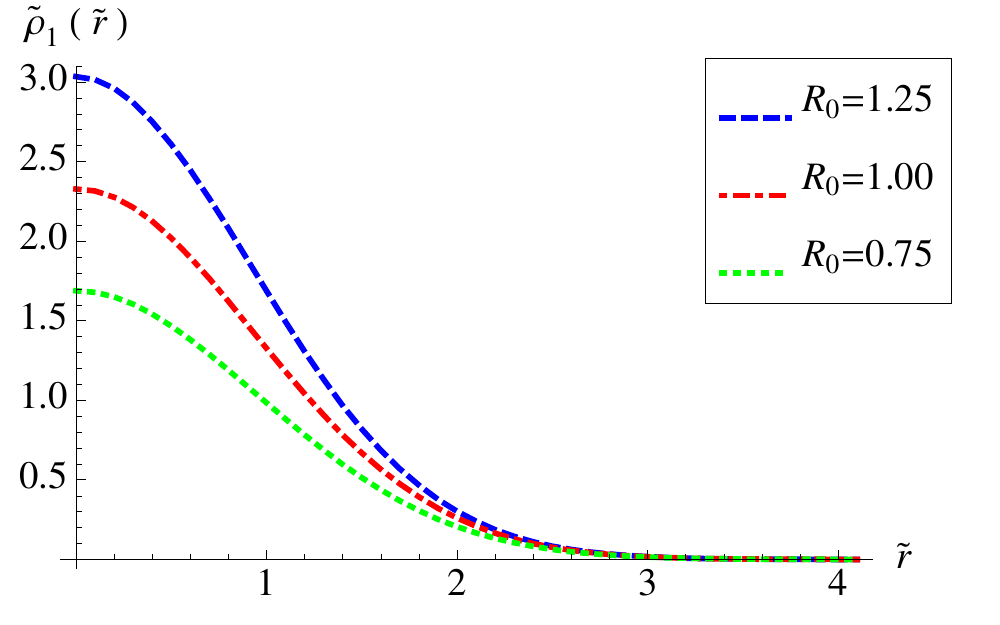}\label{fig:droplet_rho_gamma=2}} 
\subfigure[Magnetic Field for $\gamma =0.5$]{\includegraphics[width=2.25in]{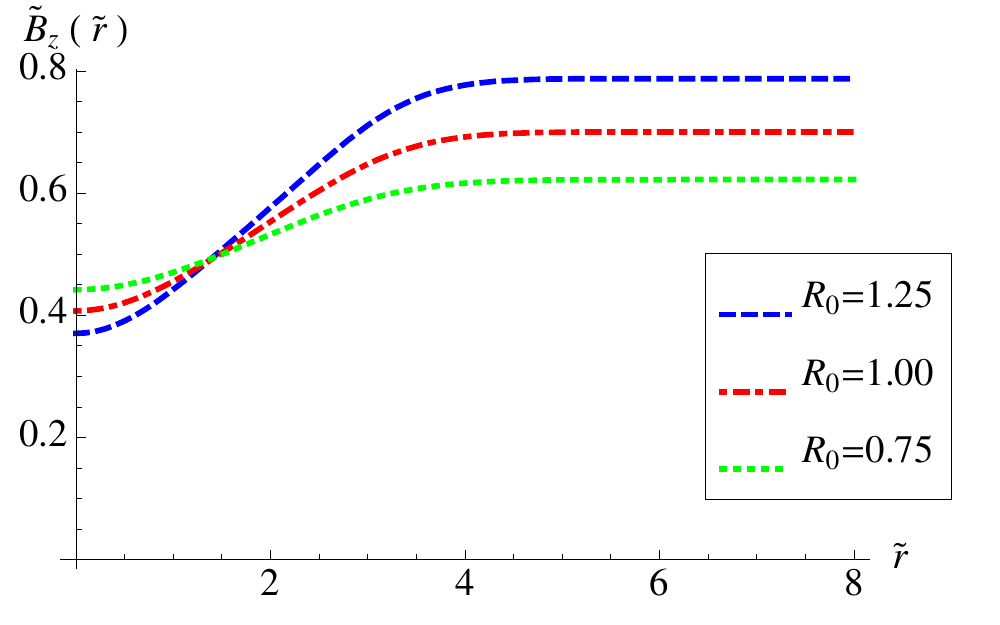}\label{fig:droplet_magnetic_gamma=0.5}} \hspace{0.5cm}
\subfigure[Magnetic for $\gamma =2$]{\includegraphics[width=2.25in]{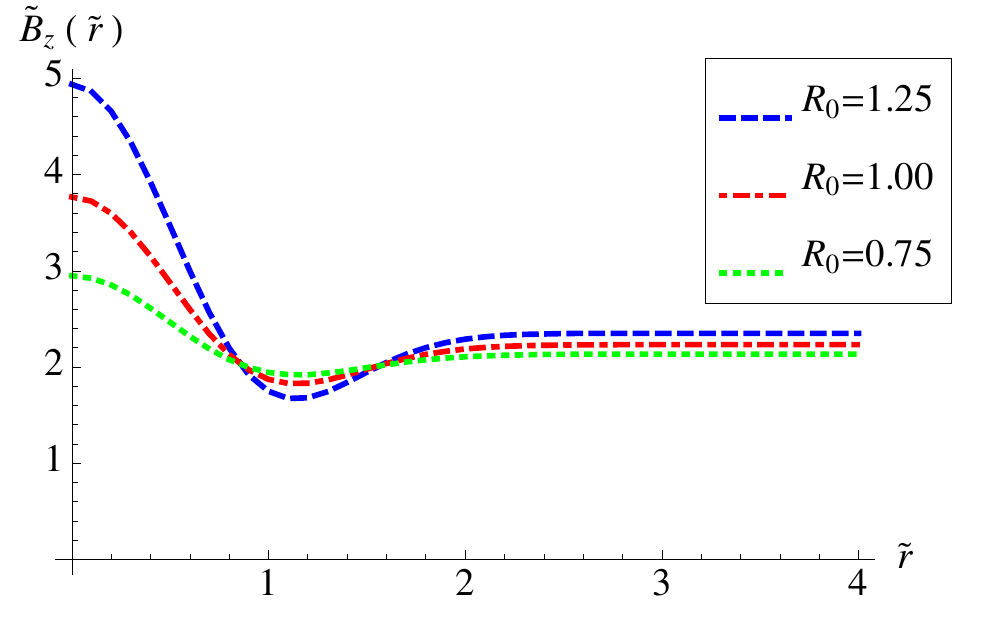}\label{fig:droplet_magnetic_gamma=2}} 
\subfigure[Current density for $\gamma =0.5$]{\includegraphics[width=2.25in]{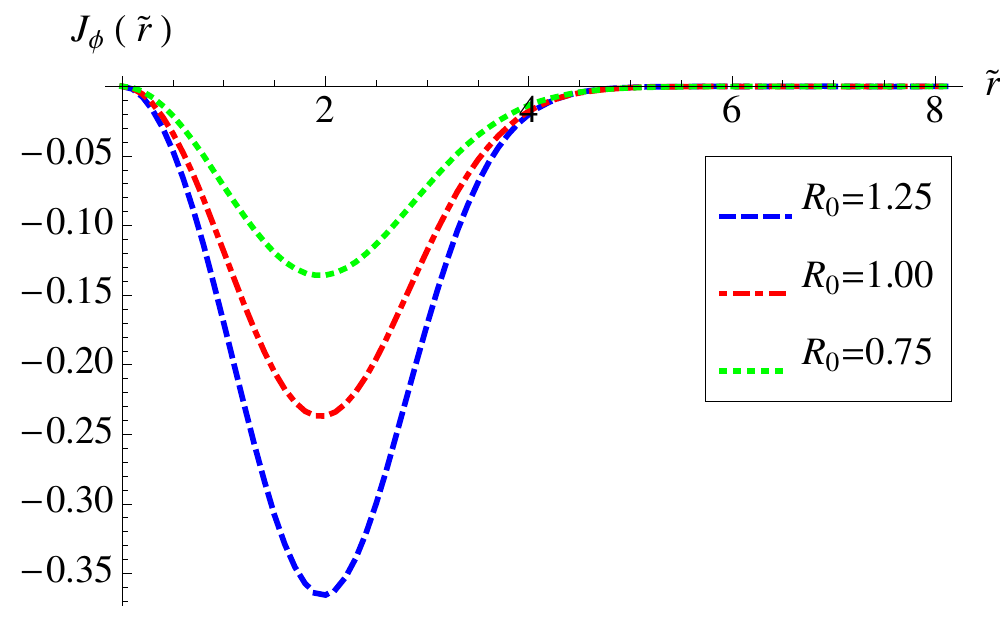}\label{fig:droplet_current_gamma=0.5}} \hspace{0.5cm}
\subfigure[Current density for $\gamma =2$]{\includegraphics[width=2.25in]{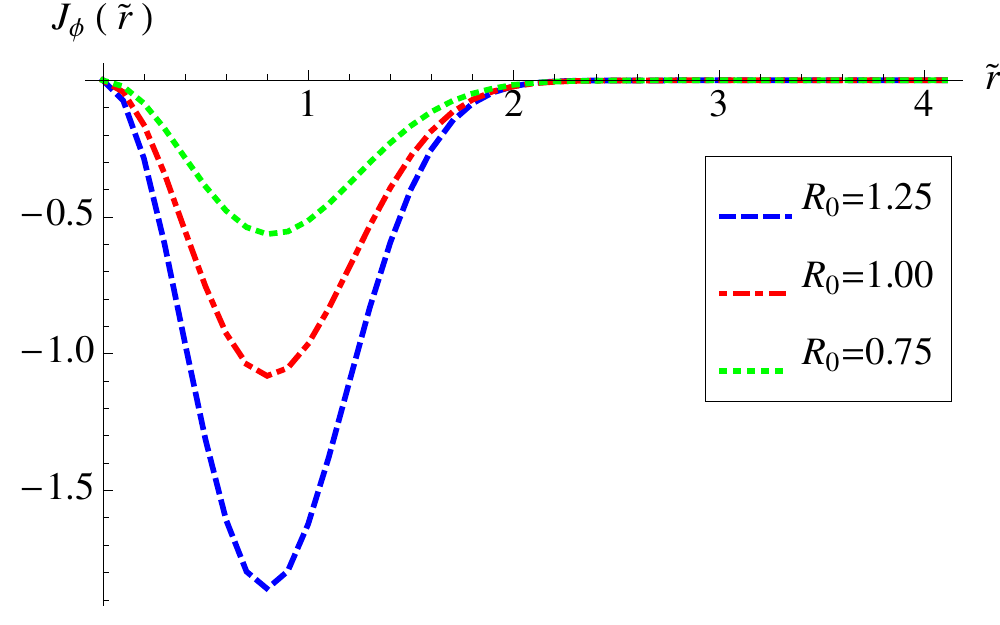}\label{fig:droplet_current_gamma=2}} 
\subfigure[Charge density for $\gamma =0.5$]{\includegraphics[width=2.25in]{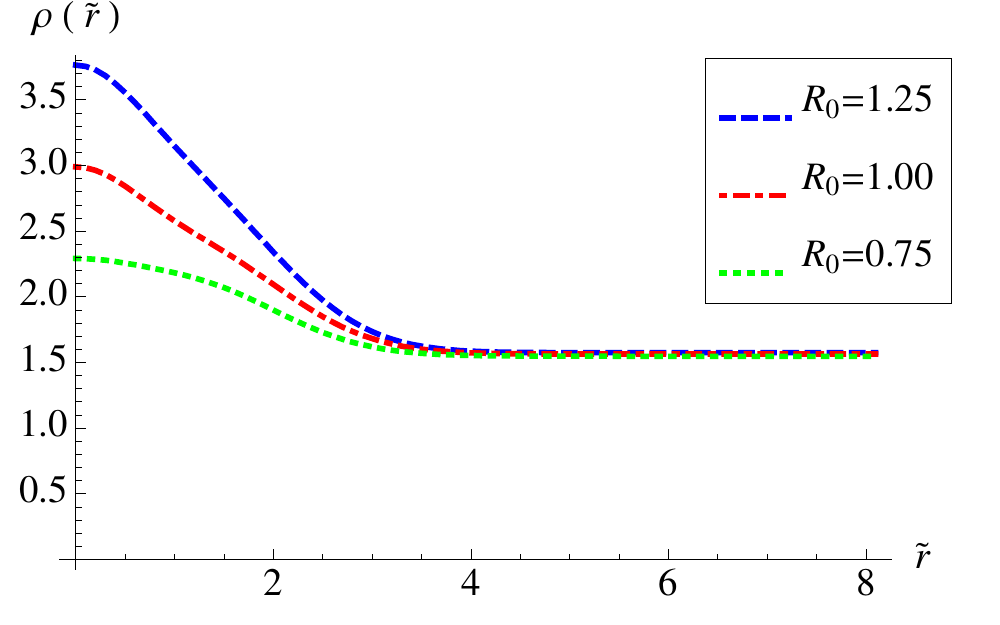}\label{fig:droplet_charge_gamma=0.5}} \hspace{0.5cm}
\subfigure[Charge density for $\gamma =2$]{\includegraphics[width=2.25in]{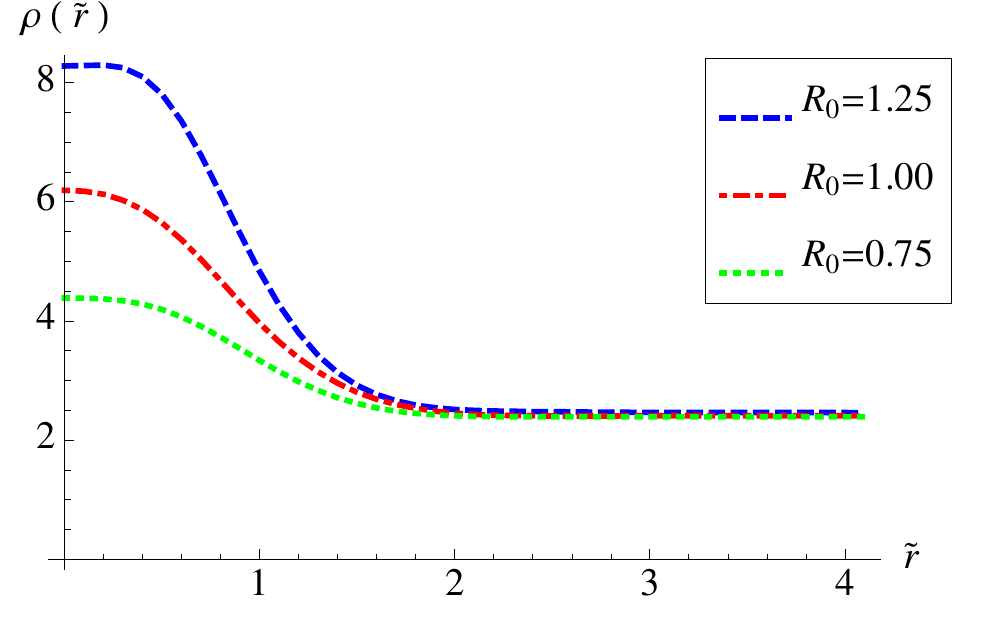}\label{fig:droplet_charge_gamma=2}} 
\caption{\small Droplet solutions for $\gamma =0.5$ on the left and $\gamma = 2$ on the right.  They correspond to $T/T_c \approx 0.84$ and $T/T_c \approx 0.67$ respectively.} \label{fig:droplet_solution}
\end{center}
\end{figure}

To give a sense of how the solutions behave, we present multiple
solutions for multiple values of $\gamma$ and $\tilde{R}_0(1)$.  The
solutions are presented in figure \ref{fig:droplet_solution}.
From figures \ref{fig:droplet_charge_gamma=0.5} and
\ref{fig:droplet_charge_gamma=2}, we see that for a given value of
$\gamma$, for various initial condition values of the scalar field, we
get the same asymptotic charge density.  We learned from the spatially
independent solution that the ratio of $T/T_c$ is determined by the
charge density, and therefore we learn that $\gamma$ fixes the value
of $T/T_c$.  In fact, as $\gamma \to 0$, we have $T/T_c \to 1$ and as
$\gamma \to \infty$, we have $T/T_c \to 0$.
From figures \ref{fig:droplet_magnetic_gamma=0.5} and
\ref{fig:droplet_magnetic_gamma=2}, we see that the magnetic field
asymptotes to a constant value, which indicates that the solutions
``live'' in a background magnetic field.  As the magnitude of the
scalar field increases, the value of this background magnetic field
rises.  Note also that how the magnetic field behaves in the core
of the droplet varies considerably between low and high temperatures.
At low temperatures (figure \ref{fig:droplet_magnetic_gamma=2}), the
magnetic field is enhanced by the droplet, whereas as high
temperatures (figure \ref{fig:droplet_magnetic_gamma=0.5}), the
droplet weakens the magnetic field in the ``core.''  This variation in
behavior suggests that these
are perhaps not superconducting droplets as was
thought\cite{Albash:2008eh,Hartnoll:2008kx}.

\begin{figure}[h] 
   \centering
   \includegraphics[width=2in]{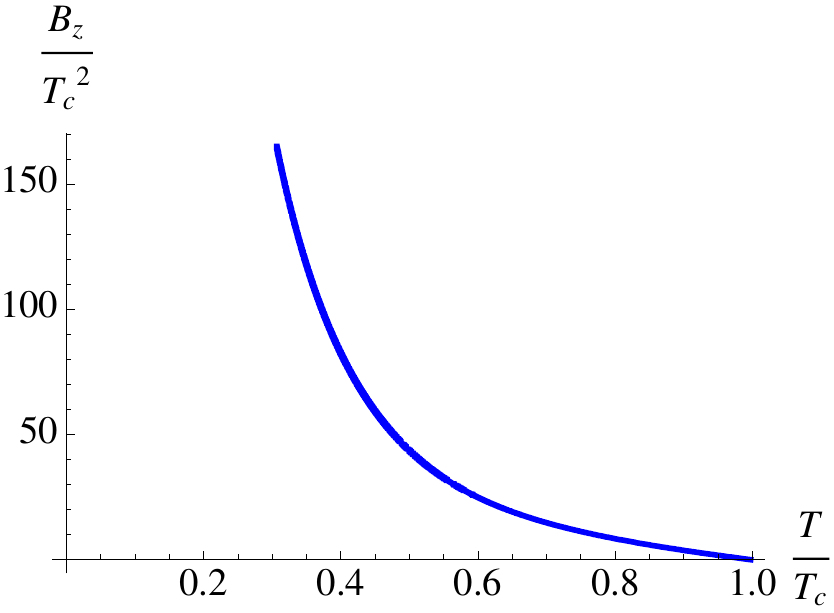} 
   \caption{\small The limiting droplet line in the $g\to\infty$
     limit. Below this, droplets disappear.}
   \label{fig:phase_diagram1}
\end{figure}

We can try to study the minimum magnetic field needed to first form
these droplet solutions.  This would correspond to studying the
problem in the limit where the magnitude of the scalar field is
approaching zero, {\it i.e.} the perturbative or probe limit.  In this
limit, it is consistent to take $\tilde{A}$ to be a constant in both
$\tilde{r}$ and $\tilde{z}$ and to take $\tilde{A}_t$ to only depend
on $z$. The equation of motion for $\tilde{R}$ with $\tilde{A} =
\gamma /2$ reduces to:
\begin{equation}
\left[\partial_{\tilde{r}}^2 \tilde{R} + \frac{1}{\tilde{r}} \partial_{\tilde{r}} \tilde{R} -\frac{1}{4} \tilde{r}^2 {\gamma}^2 \tilde{R} + \gamma \tilde{R} \right]_{z=1}  = 0 \ .
\end{equation}
This equation has (lowest energy) solution given by:
\begin{equation} \label{eqt:sol1}
\tilde{R}(\tilde{r}, z=1) = R_0(1) \exp \left(-\gamma \frac{\tilde{r}^2}{4} \right) \ .
\end{equation}
Next, we can solve for the $z$ dependence by solving:
\begin{eqnarray} \label{eqt:droplet}
& \partial_z^2 \tilde{R}  + \frac{f'}{f} \partial_z \tilde{R} - \frac{1}{f  } \gamma \tilde{R} + \frac{1}{  f^2  } \tilde{R} \tilde{A}_t^2 + \left(\frac{f'}{z f} - \frac{2}{z^2} + \frac{2}{z^2 f}\right) \tilde{R}= 0 \ , & \nonumber \\
& \partial_z^2 \tilde{A}_t = 0 \ , & 
\end{eqnarray}
with the appropriate boundary conditions at the AdS boundary.  This in
turn fixes the value of the temperature of the solution.  The value of
the magnetic field found here corresponds to the critical magnetic
field at which the droplet solutions first form.  We draw the
corresponding diagram in figure \ref{fig:phase_diagram1}.
The source of the divergence in the critical magnetic field as $T/T_c
\to 0$ is clear. As the magnetic field and charge density grow in
value, they begin to back--react on the geometry, and our decoupling
breaks down.  However, our results allow us to shed new light on the
calculation done in ref.~\cite{Albash:2008eh}. There, the prototype
solution in the droplet class was first uncovered in the probe
limit. Now we see that the probe limit is the correct limit to study
the onset of the droplets. So far in this section we've seen them in
the decoupling $g\to\infty$ limit, and further taking the probe limit
gives a limiting line at which they drop to zero height, ceasing to
exist for lower $B$.

\begin{figure}[h] 
   \centering
   \includegraphics[width=2.8in]{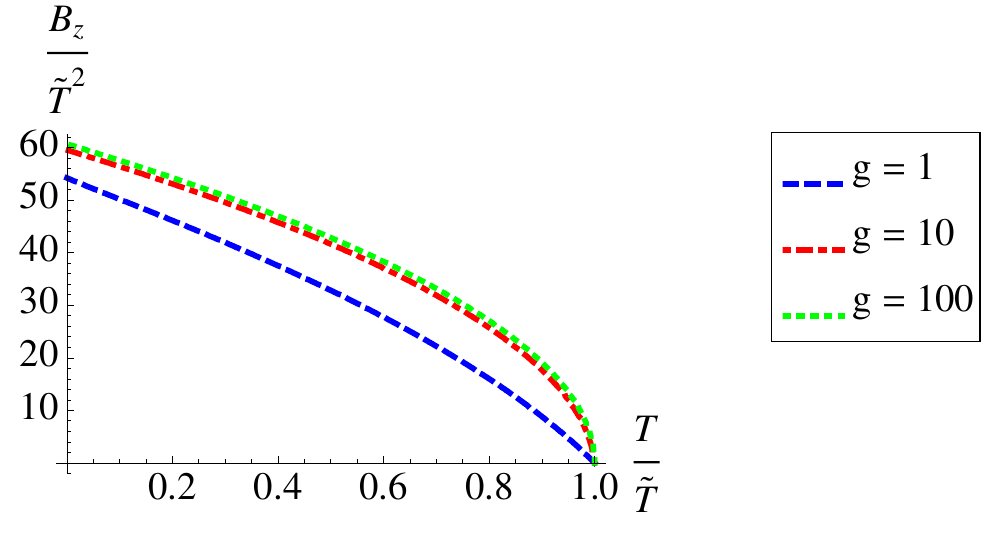} 
   \caption{\small Limiting droplet line for  three values of $g$ with $\tilde{T} = (3-q_c^2)/4 \pi$.}
   \label{fig:pd_dyonic_simple}
\end{figure}

\begin{figure}[h] 
   \centering
   \includegraphics[width=2.5in]{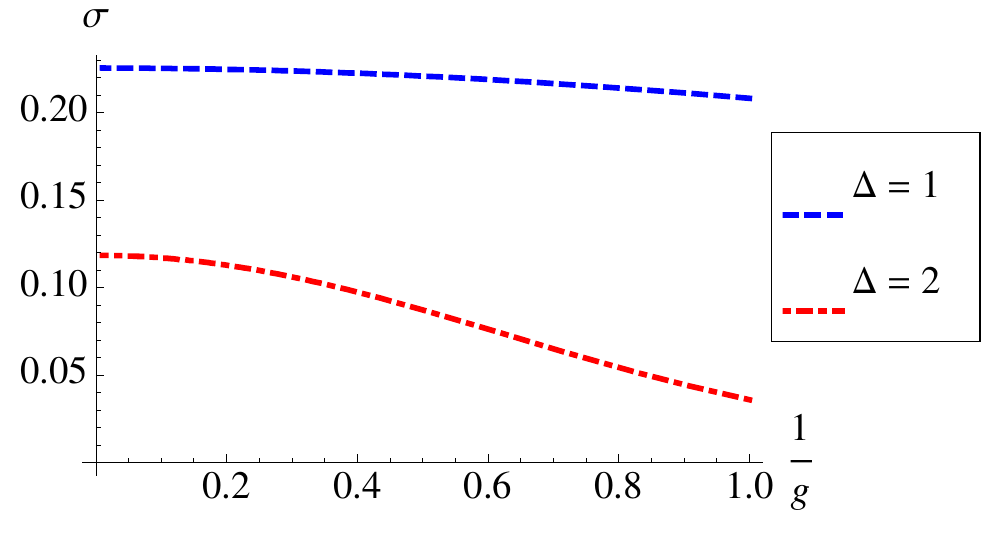} 
   \caption{\small Convergence to the $g \to \infty$ critical temperature.  The curves asymptote to 0.225492 and 0.118412 respectively. See text for discussion.}
   \label{fig:g_convergence}
\end{figure}

\begin{figure}[h] 
   \centering
   \includegraphics[width=2.8in]{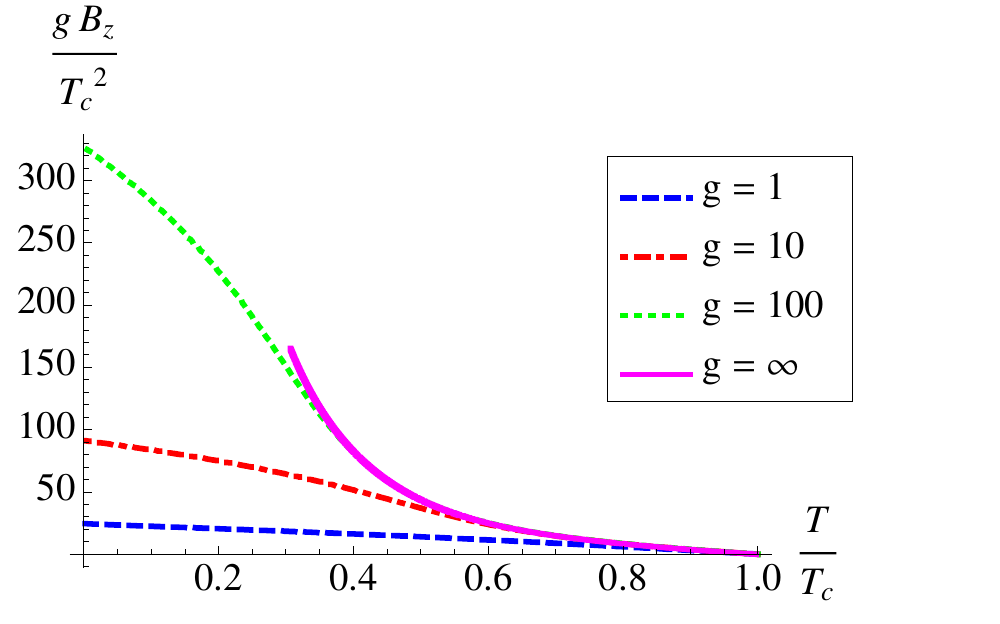} 
   \caption{\small The droplet limiting curves for a range of
     couplings, after rescaling to include the $g\to\infty$ case. $T_c
     =\alpha \sigma \sqrt{2 q g}$.}
   \label{fig:pd_unite}
\end{figure}

We can study this further (and extend to lower $T/T_c$) by working
again in the probe limit, but at arbitrary $g$, as outlined in
section~\ref{sec:probe}. Here, the background is now a charged black
hole solution, our method of taking into account some of the
back--reaction of the gauge fields. In this limit the ${\tilde\rho}$
equation of motion is given by:
\begin{equation}
\partial_z^2 \tilde{\rho} + \left( \frac{f'}{f} - \frac{2}{z} \right) \partial_z \tilde{\rho} + \frac{1}{f} \left( \partial_{\tilde{r}}^2 \tilde{\rho} + \frac{1}{\tilde{r}} \partial_{\tilde{r}} \tilde{\rho} - g^2 h^2 \tilde{r}^2 \tilde{\rho} \right) + \frac{\left(1-z\right)^2}{f^2} 4 g^2 q^2  \tilde{\rho} + \frac{2}{z^2 f} \tilde{\rho} = 0 \ ,
\end{equation}
This equation has a separable solution that we write as:
\begin{equation}
\tilde{\rho} = z Z(z) R(\tilde{r}) \ ,
\end{equation}
and we write their respective equations of motion as:
\begin{equation}
\partial_{\tilde{r}}^2 R + \frac{1}{\tilde{r}} \partial_{\tilde{r}} R - \frac{1}{4} \tilde{r}^2 \left(4 g^2 h^2 \right) R - 2 g h R = 0 \ ,
\end{equation}
\begin{equation} \label{eqt:dyonic}
\partial_z^2 Z + \frac{f'}{f}  \partial_z Z - \frac{1}{f} 2 g h Z + \frac{\left(1-z\right)^2}{f^2} 4 g^2 q^2 Z+ \left(\frac{2}{z^2 f} + \frac{f'}{f z} - \frac{2}{z^2} \right) Z= 0 \ .
\end{equation}
The solution for $R(\tilde{r})$ is given by:
\begin{equation} \label{eqt:dyonic_gaussian}
R(\tilde{r}) = \exp \left( - g h \tilde{r}^2 / 2 \right)\ .
\end{equation}
We can solve the equation for $Z(z)$ using the same arguments as before, and we get the solutions shown in figure \ref{fig:pd_dyonic_simple}.
Our claim is that the perturbative scalar field on the dyonic black
hole describes the entire phase transition line in the phase diagram,
and the problem studied earlier is simply the $g \to \infty$ limit
near $T/T_c \to 1$.  To prove this, we do several non--trivial checks.
First, we check that the dyonic theory can predict the critical
temperature of the $g \to \infty $ theory.  In order to do this, we
make the identification:
\begin{equation}
2 q g = \rho_{g \to \infty} \ ,
\end{equation}
by comparing the charge density of both theories.  Next, we know that as the magnetic field approaches zero, the droplets appear at the critical temperature (or $T/T_c = 1$) for both theories.  Given that we defined $T_c$ in different ways for both theories, by setting them equal, we should be able to calculate the relationship between $T_c$ and $\sqrt{\rho}$ that we saw in the $g \to \infty$ theory.  In particular:
\begin{equation}
T_c / \alpha = \frac{1}{4 \pi} \left( 3 - q_c^2 \right) = \sigma \sqrt{2 g q_c} \ .
\end{equation}
Therefore, we can solve for $\sigma$ as $g \to \infty$.  We present the results in figure \ref{fig:g_convergence}.  As one can see, in the limit of $g \to \infty$ we indeed recover the values 0.226 and 0.118 respectively, which were obtained earlier in the $g \to \infty$ probe case (see caption of figure \ref{fig:constant_solution}).
We can also check whether the phase diagrams coincide.  This is presented in figure \ref{fig:pd_unite}. 
As we see, the dyonic black hole results
very quickly approach our results for the $g \to \infty$ case for a
range of $T/T_c$ near one.  We are now in a position to answer what
happens when $T/T_c \to 0$ for the $g \to \infty$ limit.  The zero
temperature limit requires us to take:
 \begin{equation}
 q^2 + h^2 = 3 \ .
 \end{equation}
 As $g \to \infty$, we find that regularity of any solution at zero temperature requires us to take:
 \begin{equation}
 h \to \sqrt{3} - \frac{3}{2} \frac{1}{g} \ , \quad q  \to  3^{3/4} \frac{1}{\sqrt{g}} \ .
 \end{equation}
 Therefore, the Gaussian profile in equation
 \reef{eqt:dyonic_gaussian} vanishes in the limit of zero temperature
 and $g \to \infty$ (note that this does not happen at finite $g$).
 Therefore, the droplet no longer exists in that limit.  Another way
 to see this is that in this limit, the dimensionless quantity $g B_z
 / T_c^2$ as used in figure \ref{fig:pd_unite} diverges.
%

\subsection{Vortex Solutions}
We now consider solutions for the scalar that asymptote to a constant
non--zero value. These are the vortex solutions.
\subsubsection{Numerical Procedures}
The numerical procedure for the vortex is almost identical to that of
the droplet (see section \ref{sec:NP}). We found it much more difficult to solve for the initial
functions on the horizon using a shooting method, and so we inserted
an initial guess function for an approximation to the scalar field
$\tilde{R}( \tilde{r},1)$ at the event horizon, parameteriszed by two
constants $R_0$ and $R_1$.  We then use that function to solve for the
field $\tilde{A}(\tilde{r},1)$, however it turns that for a given
$A_0(1)$ and $R_0$, there is a specific $R_1$ that leads to a regular
solution; we again use a shooting method to determine this constant.
The constant $T_0(1)$ is determined in the same way.  With both
$\tilde{R}(\tilde{r},1)$ and $\tilde{A}(\tilde{r},1)$ determined, we
have fully fixed $\partial_z \tilde{R} (\tilde{r},1)$. The bulk
shooting problem is tackled as before. Note that the leading term in
the expansion of $\partial_z \tilde{T}$ is the constant already
determined in the spatially independent problem using
equations~\reef{eq:conditions}, since our vortices asymptote to that
case.
\subsubsection{Sample Solutions}
\begin{figure}[h!]
\begin{center}
\subfigure[Scalar Field]{\includegraphics[width=2.25in]{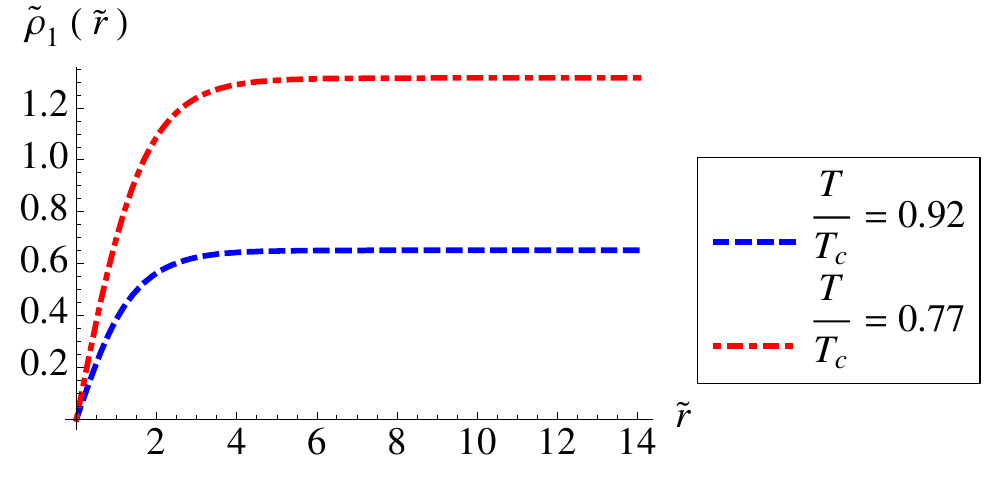}\label{fig:vortex_linear_rho1}} \hspace{1.0cm}
\subfigure[Scalar Field]{\includegraphics[width=2.25in]{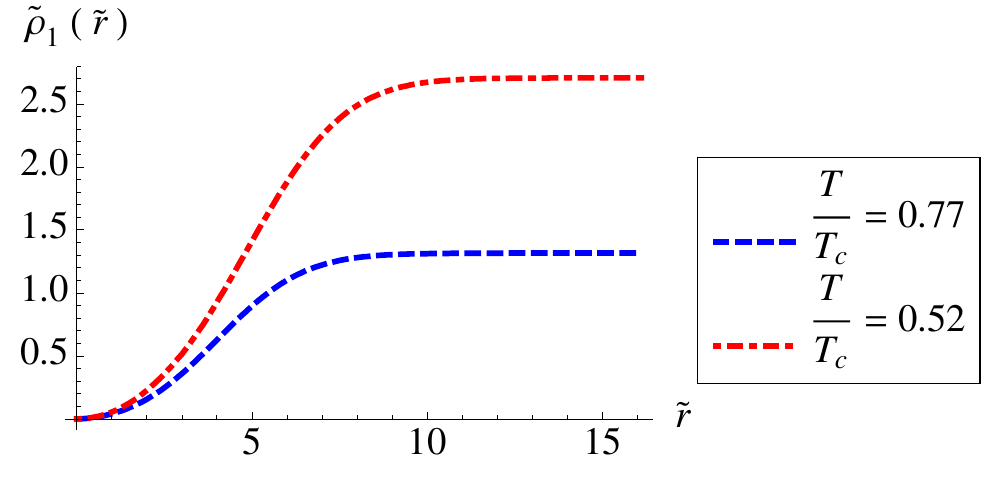}\label{fig:vortex_quad_rho1}} \hspace{1.0cm}
\caption{\small Vortex Solutions for $\xi = 1$ (LHS) and $\xi = 2$ (RHS).} \label{fig:vortex_solution1}
\end{center}
\end{figure}

\begin{figure}[h!]
\begin{center}
\subfigure[Charge Density]{\includegraphics[width=2.25in]{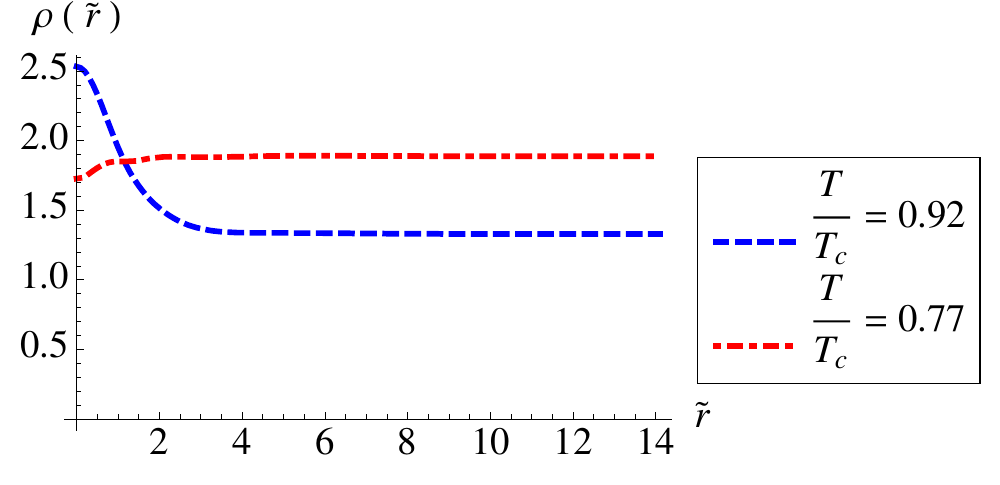}\label{fig:vortex_linear_charge}} \hspace{1.0cm}
\subfigure[Charge Density]{\includegraphics[width=2.25in]{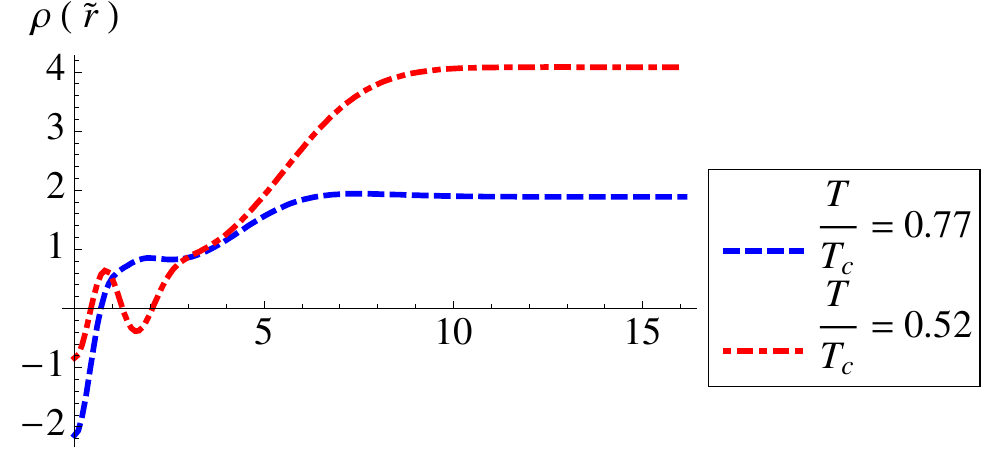}\label{fig:vortex_quad_charge}} \hspace{1.0cm}
\subfigure[Magnetic Field]{\includegraphics[width=2.25in]{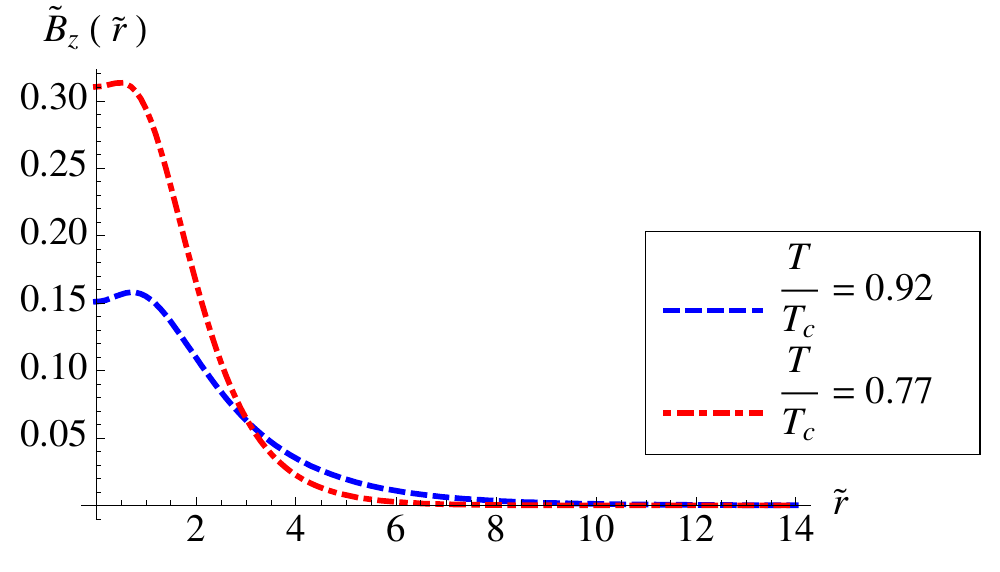}\label{fig:vortex_linear_magnetic}} \hspace{1.0cm}
\subfigure[Magnetic Field]{\includegraphics[width=2.25in]{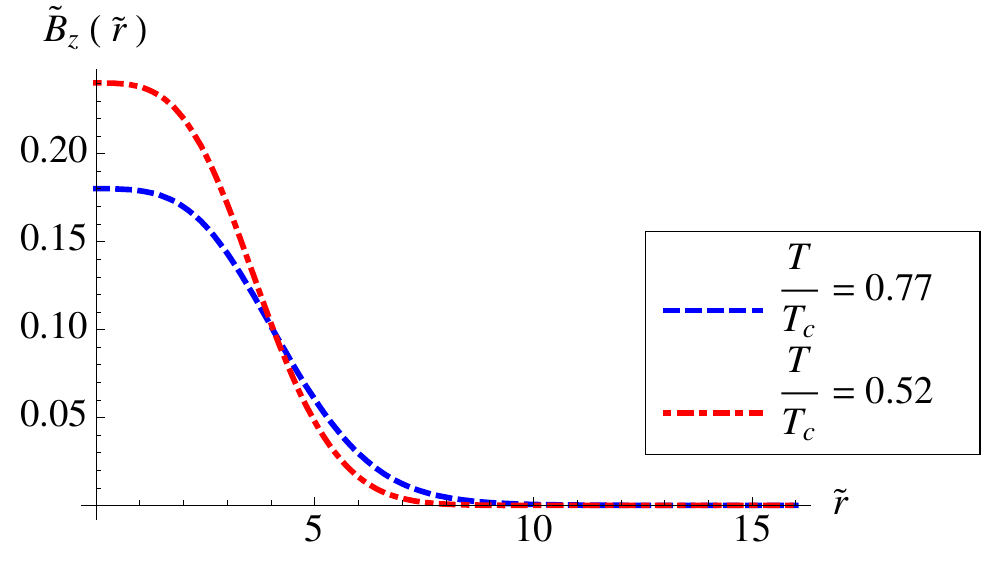}\label{fig:vortex_quad_magnetic}} \hspace{1.0cm}
\subfigure[Current density]{\includegraphics[width=2.25in]{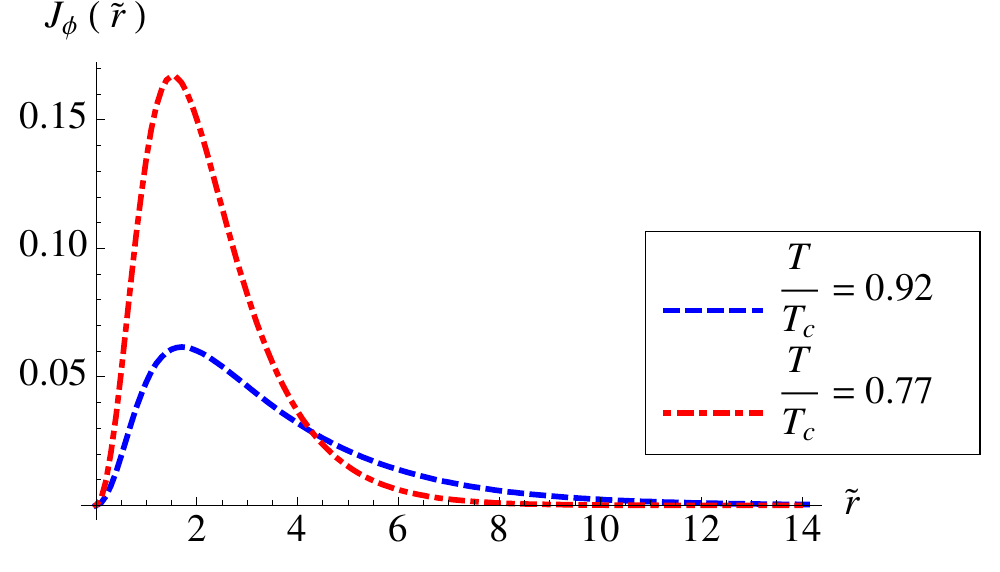}\label{fig:vortex_linear_current}} \hspace{1.0cm}
\subfigure[Current density]{\includegraphics[width=2.25in]{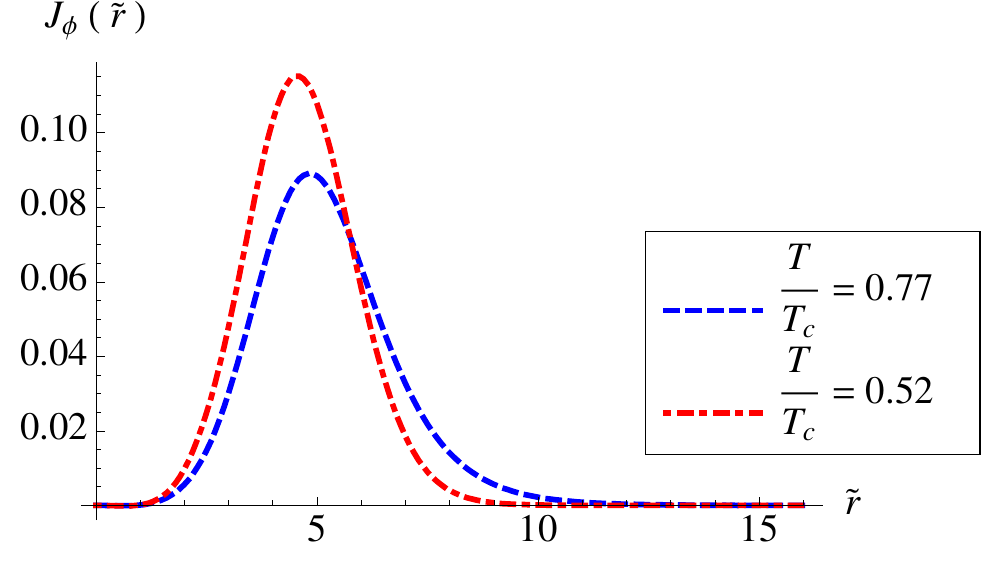}\label{fig:vortex_quad_current}} \hspace{1.0cm}
\subfigure[$A_\phi$]{\includegraphics[width=2.0in]{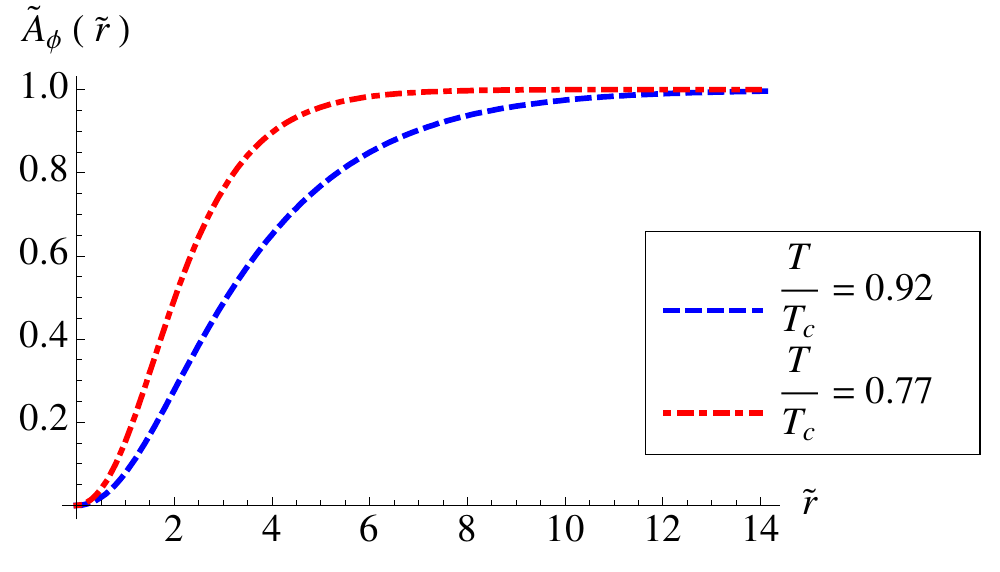}\label{fig:vortex_linear_Aphi}}\hspace{1.0cm}
\subfigure[$A_\phi$]{\includegraphics[width=2.0in]{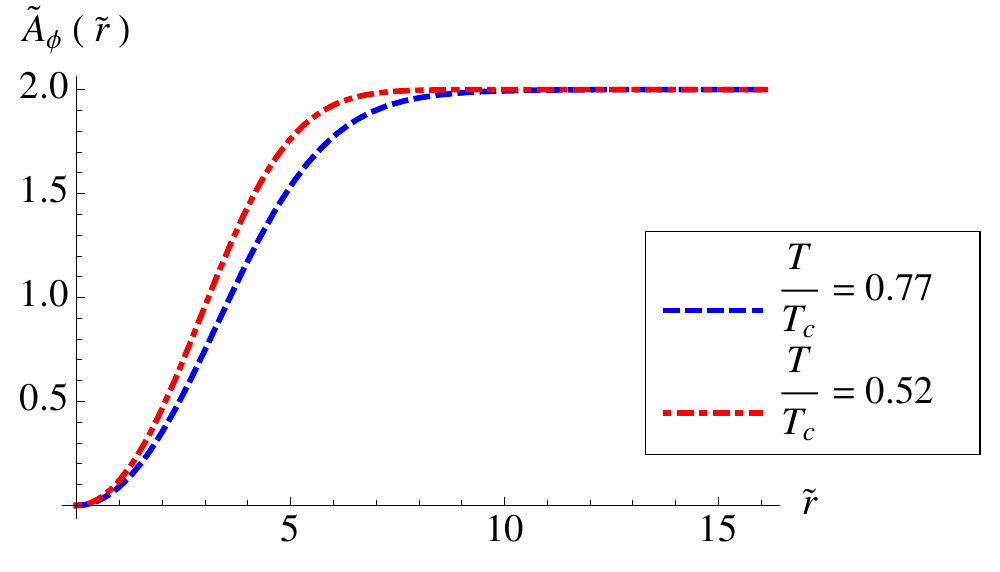}\label{fig:vortex_quad_Aphi}}
\caption{\small Vortex Solutions for $\xi = 1$ (LHS) and $\xi = 2$ (RHS).} \label{fig:vortex_solution2}
\end{center}
\end{figure}

Here, we illustrate the case of $\xi=1$ and $\xi=2$ and again use ${\cal O}_1$ as our order parameter.  We again
consider the equations of motion given in equations \reef{eqt:eom_xi}
and \reef{eqt:eom_horizon_xi}.  For simplicity, we focus on the case
of $\partial_z \tilde{A} = 0$.  We note that as the solutions approach
a constant value, they should asymptote to the spatially--independent
solutions we have presented earlier.  This in turn allows us to define
the temperature at which a given solution exists.  We present examples
of such solutions in figures \ref{fig:vortex_solution1} and
\ref{fig:vortex_solution2}.
In figure \ref{fig:vortex_solution1}, we see that very far away from
the origin, the scalar field has a constant vev, but as it approaches
the core of the vortex, it decreases to zero value.  The behavior at
the origin is of course determined by the choice of $\xi$.  
The choice of $\xi$ also influences how the gauge field
$\tilde{A}_\phi$ behaves.  In figures \ref{fig:vortex_linear_Aphi} and
\ref{fig:vortex_quad_Aphi}, we see that the value of $\tilde{A}_\phi$
asymptotes to the value of $\xi$.  This is exactly the behavior
required for the magnetic flux penetrating the vortex to be quantized
with value $2 \pi \xi$.  We review this briefly in
Appendix~\ref{sec:flux_quantization}. Indeed, $\xi$ defines a
non--trivial topological winding number: The scalar ${\tilde\rho}$
becomes constant at infinity, breaking the $U(1)$. Therefore the gauge
symmetry of equation~\reef{eq:gauge} cannot be used to unwind
$\theta$. Gauge symmetry is unbroken at infinity for the droplets, so
$\xi$ is not a winding number for them.  The current density $J_\phi({\tilde
  r})$ (figures \ref{fig:vortex_linear_current} and \ref{fig:vortex_quad_current}) is zero asymptotically and peaks in a ring around the core, supporting the magnetic field, as expected for
a vortex.

In figure \ref{fig:vortex_quad_charge},
we find that the charge density near the origin begins to oscillate as
the density drifts downwards.  It is difficult to say whether or not
this is a physical attribute of the solutions or whether it is an
artifact of our scheme to find the appropriate shooting functions that
satisfy the $z=0$ boundary condition.  If they are physical, they may be
caused by screening effects being strong in the core of the vortex.
It is interesting to note that this behavior appears to be absent for
the $\xi = 1$ vortex (see figure \ref{fig:vortex_linear_charge}),
although we do see that there is a transition from the charge density
increasing in the core to decreasing in the core as the temperature is
lowered.  Another curiosity for the $\xi = 1$ vortex is the behavior
of the magnetic field near the origin (see figure
\ref{fig:vortex_linear_magnetic}.  Instead of flattening out as is the
case for the $\xi = 2$ solutions, it dips slightly downwards.  We expect this also to be a numerical artifact, since the $\xi = 1$ case is more numerically sensitive because of its sharper profile near $\tilde{r} = 0$.
%

\section{Stability and Quasinormal Modes}
\label{stability}
In this section we consider the stability of our solutions to
fluctuations in the fields. We work in the decoupling limit and
proceed by considering fluctuations about our classical solutions:
\begin{equation}
  A_\mu = A_\mu^{(0)} + A_\mu^{(1)} \ , \quad {\tilde\rho} = {\tilde\rho}^{(0)} + {\tilde\rho}^{(1)} \ , \quad \theta = \theta^{(0)} + \theta^{(1)} \ ,
\end{equation}
where the fields with superscripts~$(0)$ are the background fields and
the fields with superscipts~$(1)$ are the fluctuation fields.  Note in
particular that because we are working in the $g \to \infty$ limit, we
do not consider fluctuations of the metric since those come at
$O(1/g^2)$.  The action to quadratic order in the fluctuations is
given by:
\begin{eqnarray}
S&=&\frac{1}{2 g^2 \kappa_4^2} \int d^4 x \sqrt{-G} \left( -\frac{1}{2} G^{\mu \nu} \left[\partial_\mu \tilde{\rho}^{(1)} \partial_\nu \tilde{\rho}^{(1)} + \left(\tilde{\rho}^{(1)}\right)^2 \left( A^{(0)}_\mu - \partial_\mu \theta^{(0)} \right) \left( A^{(0)}_\nu - \partial_\nu \theta^{(0)} \right) \right. \right. \nonumber \\
&& \left. \left.+4 \tilde{\rho}^{(0)} \tilde{\rho}^{(1)} \left( A^{(0)}_\mu - \partial_\mu \theta^{(0)} \right) \left( A^{(1)}_\nu - \partial_\nu \theta^{(1)} \right) + \left(\tilde{\rho}^{(0)}\right)^2 \left( A^{(1)}_\mu - \partial_\mu \theta^{(1)} \right) \left( A^{(1)}_\nu - \partial_\nu \theta^{(1)} \right) \right] \right. \nonumber \\
&&\left. + \frac{1}{L^2} \left(\tilde{\rho}^{(1)}\right)^2 - \frac{L^2}{4} F^{(1)}_{\mu \nu} F^{\mu \nu}_{(1)} \right) \ .
\end{eqnarray}
The resulting equations of motion are given by:
\begin{eqnarray}
& \frac{1}{\sqrt{-G}} \partial_\mu \left(\sqrt{-G} G^{\mu \nu} \partial_\nu \tilde{\rho}^{(1)} \right) - \tilde{\rho}^{(1)} G^{\mu \nu} \left( A^{(0)}_\mu - \partial_\mu \theta^{(0)} \right) \left( A^{(0)}_\nu - \partial_\nu \theta^{(0)} \right) & \nonumber  \\ 
&\hskip4.5cm- 2 \tilde{\rho}^{(0)} G^{\mu \nu} \left( A^{(0)}_\mu - \partial_\mu \theta^{(0)} \right)\left(A^{(1)}_\nu - \partial_\nu \theta^{(1)} \right) + \frac{2}{L^2} \tilde{\rho}^{(1)} = 0\ , & \nonumber  \\
&\frac{1}{\sqrt{-G}} \partial_\mu \left(\sqrt{-G} \left(2 \tilde{\rho}^{(0)} \tilde{\rho}^{(1)} G^{\mu \nu} \left(A^{(0)}_\nu - \partial_\nu \theta^{(0)} \right) + \left(\tilde{\rho}^{(0)}\right)^2 G^{\mu \nu} \left( A^{(1)}_\nu - \partial_\nu \theta^{(1)} \right) \right) \right) = 0\ , & \nonumber  \\
&\frac{L^2}{\sqrt{-G}} \partial_\nu \left(\sqrt{-G} G^{\nu \lambda} G^{\mu \sigma} \hat{F}_{\lambda \sigma} \right) - 2 G^{\mu \nu} \tilde{\rho}^{(0)} \tilde{\rho}^{(1)} \left(A^{(0)}_\nu - \partial_\nu \theta^{(0)} \right) &\nonumber\\
&\hskip8.5cm - \left(\tilde{\rho}^{(0)}\right)^2 G^{\mu \nu} \left(A^{(1)}_\nu - \partial_\nu \theta^{(1)} \right) = 0\ . & \nonumber
\end{eqnarray}
\subsection{Spatially Independent Solution}
We can consider the ansatz:
\begin{equation}
A_x^{(1)} = \alpha e^{- i \omega t} \tilde{A}_x^{(1)} (\tilde{\omega}, z)\ .
\end{equation}
The equation of motion for the fluctuation field is given by:
\begin{equation}
\partial_z^2 \tilde{A}_x^{(1)} + \frac{f'}{f} \partial_z \tilde{A}_x^{(1)} + \frac{\tilde{\omega}^2}{f^2} \tilde{A}_x^{(1)} - \frac{1}{f} \left(\tilde{R}^{(0)} \right)^2 \tilde{A}_x^{(1)}  = 0\ .
\end{equation}
Near the event horizon, the field satisfies an equation of the form:
\begin{equation}
\partial_z^2 \tilde{A}_x^{(1)} - \frac{1}{1-z} \partial_z \tilde{A}_x^{(1)} + \frac{\tilde{\omega}^2}{9 (1 - z)^2 } \tilde{A}_x^{(1)} = 0 \ .
\end{equation}
This has solutions given by ingoing (negative) and outgoing (positive) waves:
\begin{equation}
\tilde{A}_x^{(1)} \propto \left( 1- z \right)^{\pm i \tilde{\omega}/3} \ .
\end{equation}
The appropriate condition to have at the event horizon is of ingoing
waves, and therefore, we make the following field redefinition:
\begin{equation}
\tilde{A}_x^{(1)} = \left(1 - z\right)^{-i \tilde{\omega}/3} \chi (\tilde{\omega}, z) \ .
\end{equation}
The equation of motion is now given by:
\begin{eqnarray}
&&\partial_z^2 \chi - \frac{\tilde{\omega}^2}{9} \frac{1}{\left(1-z \right)^2} \chi + \frac{f'}{f}  \partial_z \chi + \frac{\tilde{\omega}^2}{f^2} \chi - \frac{1}{f} \left( \tilde{R}^{(0)} \right)^2 \chi \nonumber\\
&&\hskip5cm+ \frac{i \tilde{\omega}}{3} \left( \frac{2}{1-z} \partial_z \chi + \frac{1}{\left(1-z \right)^2} \chi  + \frac{f'}{f} \frac{1}{1-z} \chi \right) = 0\ . 
\end{eqnarray}
Expanding near the event horizon, we get the following restrictions on
the initial conditions:
\begin{equation}
\left[ \partial_z \chi =\frac{3 i \tilde{\omega} + 2 \tilde{\omega}^2 - 3 \left(\tilde{R}^{(0)}\right)^2}{9 -6 i \tilde{\omega} } \chi \right]_{z=1} \ .
\end{equation}
By requiring that $\chi(z=0) = 0$, we find that this condition is only
satisfied for discrete values of $\tilde{\omega}$.  We present some of
the values in table \ref{table:complex_omega1}.  In particular, since
the imaginary part of $\tilde{\omega}$ is negative, this corresponds
to having fluctuations that decay away.  Therefore, the constant
solutions are stable under fluctuations.
\begin{table}[!h] 
\begin{center}
\begin{tabular}{|c|c|} 
\hline $n$ & $\tilde{\omega}$ \\
\hline 0 & 1.42804 -  1.87689 $i$ \\
\hline 1 & 3.07885 -  2.79650 $i$ \\
\hline
\end{tabular}
\caption{Values of $\tilde{\omega}$ that give regularizable solutions at $T/T_c = 0.772$.} \label{table:complex_omega1}
\end{center} 
\end{table}
%
\subsubsection{Hydrodynamic Limit and Conductivity}
%
As an aside, we can push our analysis a little more to compute the DC
conductivity (already done in ref.\cite{Hartnoll:2008vx}). We proceed
by studying the problem in the limit where $\tilde{\omega}$ approaches
zero.  We therefore consider solving for the field
$\chi(\tilde{\omega},z)$ as an expansion in $\tilde{\omega}$:
\begin{equation}
\chi (\tilde{\omega},z) = \chi_0 (z) + \tilde{\omega} \chi_1 (z) + \dots
\end{equation} 
As explained in ref.~\cite{Kovtun:2005ev}, in the hydrodynamic limit
and where the field is solved as an expansion in $\tilde{\omega}$, the
normalizibility condition cannot be satisfied by the terms $\chi_i$.
We proceed by focusing on the $\chi_0$ term.  The equation of motion
is given by:
\begin{equation}
\partial_z^2 \chi_0 + \frac{f'}{f} \partial_z \chi_0 - \frac{1}{f} \left(\tilde{R}^{(0)} \right)^2 \chi_0 = 0\ .
\end{equation}
Note that at the AdS ($z=0$) boundary the solution for $\chi_0$ is
given by:
\begin{equation}
\lim_{z \to 0} \chi_0  = \left\{ \begin{array}{cr}
a_x & ; \ T > T_c \ , \\
a_x + j_x z & ; \  T < T_c \ ,
\end{array} \right.
\end{equation}
where $a_x$ and $j_x$ are constants and $j_x$ is proportional to the
current.  In particular, if we define the conductivity using:
\begin{equation}
\sigma  = \frac{J_x}{E_x} = \frac{J_x}{ i \omega A_x}\ .
\end{equation}
Therefore for $T<T_c$, $\mathrm{Im} (\sigma ) \propto
\tilde{\omega}^{-1}$, and therefore by the Kramers--Kronig relations,
we have that $\mathrm{Re}(\sigma ) \propto \delta (\tilde{\omega})$ as
shown in ref.~\cite{Hartnoll:2008vx}.
%
\subsection{Spatially Dependent Solutions}
\begin{figure}[h!]
\begin{center}
\subfigure[Droplet Fluctuation with $\tilde{\omega} = 6.06 - 2.41 i$ and $T/T_c = 0.85$.]{\includegraphics[width=2.25in]{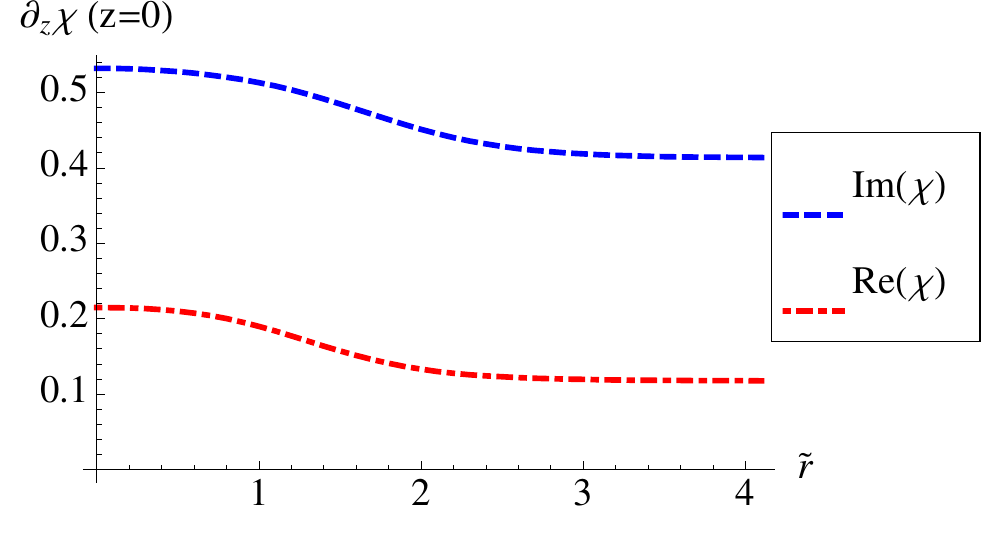}\label{fig:droplet_fluctuation}} \hspace{1.0cm}
\subfigure[Linear Vortex Fluctuation with $\tilde{\omega} = 13.12 - 2.32 i$ and $T/T_c = 0.77$.]{\includegraphics[width=2.25in]{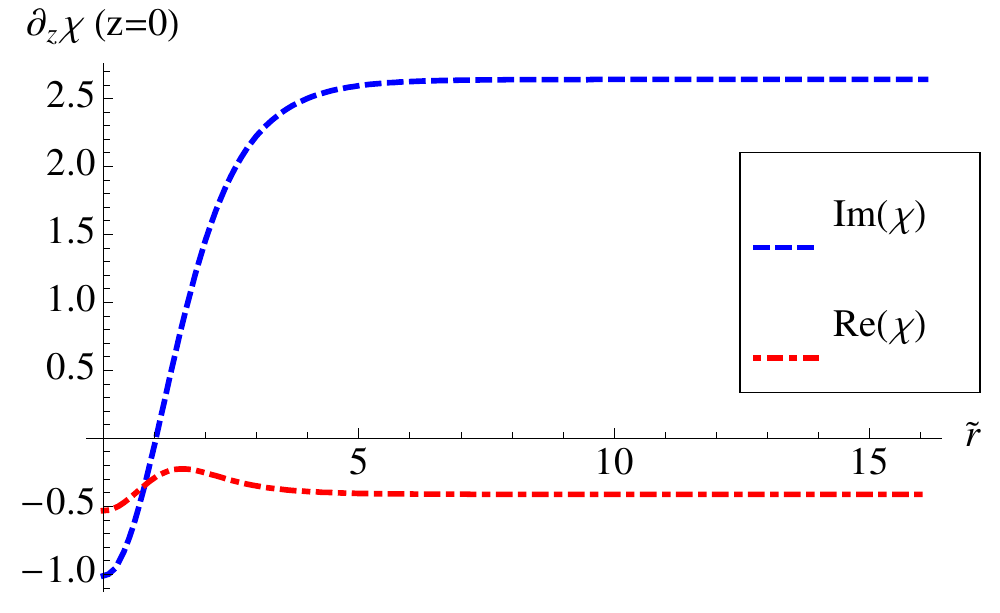}\label{fig:vortex_fluctuation}}
\caption{\small Solutions for the fluctuation field $\chi$.} \label{fig:fluctuations}
\end{center}
\end{figure}

We begin by slightly simplifying by considering the following field
behavior:
\begin{eqnarray}
& \rho^{(1)} =  e^{-i \omega t} \tilde{\rho}^{(1)} (\tilde{\omega}, \tilde{r}, z) \ , \quad A_t^{(1)} =  \alpha e^{-i \omega t} \tilde{A}_t^{(1)} (\tilde{\omega}, \tilde{r}, z)  \ , \quad A_\phi^{(1)} = e^{-i \omega t} \tilde{A}_\phi^{(1)} (\tilde{\omega}, \tilde{r}, z)\ ,& \\
& A_r^{(1)} = \alpha e^{-i \omega t} \tilde{A}_r^{(1)} (\tilde{\omega}, \tilde{r}, z) \ , \quad  A_z^{(1)} =  \alpha e^{-i \omega t} \tilde{A}_z^{(1)} (\tilde{\omega}, \tilde{r}, z) \ , \quad \theta^{(1)} = 0 \ . &
\end{eqnarray}
where we have defined dimensionless fields and variables $\omega = \alpha \tilde{\omega}$.  This reduces the equations to:
\begin{eqnarray}
&&\partial_{\tilde{r}}^2 \tilde{A}_z^{(1)} - \partial_{\tilde{r}} \partial_z \tilde{A}_r^{(1)} + \frac{1}{\tilde{r}} \left( \partial_{\tilde{r}} \tilde{A}_z^{(1)} - \partial_z \tilde{A}_r^{(1)} \right) - \frac{1}{f} \left( - \tilde{\omega}^2 \tilde{A}_z^{(1)} + i \tilde{\omega} \partial_z \tilde{A}_t^{(1)} \right) \nonumber\\
&&\hskip9cm- \frac{1}{z^2} \left(\tilde{\rho}^{(0)}\right)^2 \tilde{A}_z^{(1)} = 0 \ , \nonumber \\
&&\partial_z^2 \tilde{A}_r^{(1)}  -   \partial_{\tilde{r}} \partial_z  \tilde{A}_z^{(1)} + \frac{f'}{f} \left( \partial_z \tilde{A}_r^{(1)} - \partial_{\tilde{r}} \tilde{A}_z^{(1)} \right) - \frac{1}{f^2} \left( - \tilde{\omega}^2 \tilde{A}_r^{(1)} + i \tilde{\omega} \partial_{\tilde{r}} \tilde{A}_t^{(1)} \right) \nonumber\\
&&\hskip9cm- \frac{1}{z^2 f} \left( \tilde{\rho}^{(0)}\right)^2 \tilde{A}_r^{(1)} = 0\ ,  
\end{eqnarray}
and
\begin{eqnarray}
&& \partial_z^2 \tilde{A}_t^{(1)} + i \tilde{\omega} \partial_z \tilde{A}_z^{(1)} + \frac{1}{f} \left( \partial_{\tilde{r}}^2 \tilde{A}_t^{(1)} + i \tilde{\omega} \partial_{\tilde{r}} \tilde{A}_r^{(1)} + \frac{1}{\tilde{r}} \left( \partial_{\tilde{r}} \tilde{A}_t^{(1)} + i \tilde{\omega} \tilde{A}_r^{(1)} \right) \right) - \frac{2}{z^2 f} \tilde{\rho}^{(0)} \tilde{\rho}^{(1)} \tilde{A}_t^{(0)} \nonumber\\
&&\hskip11cm- \frac{1}{z^2 f} \left(\tilde{\rho}^{(0)} \right)^2 \tilde{A}_t^{(1)}= 0 \ ,\nonumber \\
&&\partial_z^2 \tilde{A}_\phi^{(1)} + \frac{f'}{f} \partial_z \tilde{A}_\phi^{(1)} + \frac{1}{f} \left( \partial_{\tilde{r}}^2 \tilde{A}_\phi - \frac{1}{\tilde{r}} \partial_{\tilde{r}} \tilde{A}_\phi^{(1)}  \right) +\frac{1}{f^2} \tilde{\omega}^2 \tilde{A}_\phi^{(1)} - \frac{2}{z^2 f} \tilde{\rho}^{(0)} \tilde{\rho}^{(1)} \left( \tilde{A}_\phi^{(0)} - \xi \right) \nonumber\\ 
&&\hskip11cm- \frac{1}{z^2 f} \left( \tilde{\rho}^{(0)} \right)^2 \tilde{A}_\phi^{(1)} = 0 \ , \nonumber \\
&&\partial_z^2 \tilde{\rho}^{(1)} + \left( \frac{f'}{f} - \frac{2}{z} \right) \partial_z \tilde{\rho}^{(1)} \nonumber\\
&&\hskip2cm+ \frac{1}{f} \left( \partial_{\tilde{r}}^2 \tilde{\rho}^{(1)} + \frac{1}{\tilde{r}} \partial_{\tilde{r}} \tilde{\rho}^{(1)}  - \frac{1}{\tilde{r}^2} \left( \tilde{A}_\phi^{(0)} -\xi \right)^2 \tilde{\rho}^{(1)} - \frac{2}{\tilde{r}^2} \left( \tilde{A}_\phi^{(0)} - \xi \right) \tilde{A}_\phi^{(1)} \tilde{\rho}^{(0)}\right) \nonumber \\
&& \hskip3.5cm+ \frac{1}{f^2} \tilde{\omega}^2 \tilde{\rho}^{(1)} + \frac{1}{f^2} \tilde{\rho}^{(1)} \left(\tilde{A}_t^{(0)} \right)^2 + \frac{2}{f^2} \tilde{A}_t^{(0)} \tilde{A}_t^{(1)} \tilde{\rho}^{(0)} + \frac{2}{z^2 f} \tilde{\rho}^{(1)} = 0\ .
\end{eqnarray}
Consider  two possible ans\"atze.
First, let us consider the radial gauge choice $\tilde{A}_z^{(1)} = 0 $.  In this gauge, the equation for $\tilde{A}_z^{(1)}$ gives the following restriction:
\begin{equation} \label{eqt:ansatz1_restriction}
\partial_z \left( \partial_{\tilde{r}} \tilde{A}_r^{(1)} + \frac{1}{\tilde{r}} \tilde{A}_r^{(1)} \right)  = -\frac{i \tilde{\omega}}{f} \partial_z \tilde{A}_t^{(1)} \ .
\end{equation}
This result is interesting because it suggests that in the limit of
$\tilde{\omega} \to 0$, there is a consistent solution with
$\tilde{A}_r^{(1)} = 0$.  This suggests that for this ansatz,
$\tilde{A}_r^{(1)} \propto \tilde{\omega}$.  However, implementing the
restriction in equation \reef{eqt:ansatz1_restriction} is not trivial,
and therefore, we consider a different ansatz. For lack of a better
name, we call the temporal gauge, $\tilde{A}_t^{(1)} = 0$.
In this gauge, the equation for $\tilde{A}_t^{(1)}$ gives the following restriction:
\begin{equation}
 i \tilde{\omega} \left( \partial_z \tilde{A}_z^{(1)} + \frac{1}{f} \left( \partial_{\tilde{r}} \tilde{A}_r^{(1)} + \frac{1}{\tilde{r}} \tilde{A}_r^{(1)} \right)  \right) = \frac{2}{z^2 f} \tilde{\rho}^{(1)} \tilde{\rho}^{(0)} \tilde{A}_t^{(0)} 
\ .\end{equation}
In particular, for this ansatz, in the limit of $\tilde{\omega} \to
0$, we see that we must have $\tilde{\rho}^{(1)} \propto
\tilde{\omega}$ which in turn means that we must have
$\tilde{A}_\phi^{(1)} \propto \tilde{\omega}$ for consistency.  This
particular restriction is more straightforward to implement since we
can directly insert it into the equations for $\tilde{A}_z^{(1)}$ and
$\tilde{A}_r^{(1)}$.  The resulting equations of motion are given by:
\begin{eqnarray}
&\partial_z^2 \tilde{A}_z^{(1)}  + \frac{f'}{f} \partial_z \tilde{A}_z^{(1)}  - \frac{2}{i \omega f} \partial_z \left( \frac{1}{z^2} \tilde{\rho}^{(0)} \tilde{\rho}^{(1)} \tilde{A}_t^{(0)} \right) + \frac{1}{f} \left(\partial_{\tilde{r}}^2 \tilde{A}_z^{(1)} + \frac{1}{\tilde{r}} \partial_{\tilde{r}} \tilde{A}_z^{(1)} \right) & \nonumber\\
& \hskip8cm+ \frac{\tilde{\omega}^2  }{f^2}\tilde{A}_z^{(1)}  - \frac{1}{f z^2} \left(\tilde{\rho}^{(0)}\right)^2 \tilde{A}_z^{(1)} = 0\ , & \nonumber \\
&\partial_z^2 \tilde{A}_r^{(1)}  + \frac{f'}{f}\partial_z \tilde{A}_r^{(1)} + \frac{1}{f} \left( \partial_{\tilde{r}}^2 \tilde{A}_r^{(1)} + \frac{1}{\tilde{r}} \partial_{\tilde{r}} \tilde{A}_r^{(1)} - \frac{1}{\tilde{r}^2} \tilde{A}_r^{(1)} \right)- \frac{f'}{f} \partial_{\tilde{r}} \tilde{A}_z^{(1)} - \frac{2}{z^2 f i \tilde{\omega}} \partial_{\tilde{r}} \left( \tilde{\rho}^{(0)} \tilde{\rho}^{(1)} \tilde{A}_t^{(0)} \right) & \nonumber\\
&\hskip8cm+ \frac{\tilde{\omega}^2}{f^2}  \tilde{A}_r^{(1)} - \frac{1}{z^2 f} \left( \tilde{\rho}^{(0)}\right)^2 \tilde{A}_r^{(1)} = 0\ , & \nonumber \\
&\partial_z^2 \tilde{A}_\phi^{(1)} + \frac{f'}{f} \partial_z \tilde{A}_\phi^{(1)} + \frac{1}{f} \left( \partial_{\tilde{r}}^2 \tilde{A}_\phi - \frac{1}{\tilde{r}} \partial_{\tilde{r}} \tilde{A}_\phi^{(1)}  \right) +\frac{1}{f^2} \tilde{\omega}^2 \tilde{A}_\phi^{(1)} - \frac{2}{z^2 f} \tilde{\rho}^{(0)} \tilde{\rho}^{(1)} \left( \tilde{A}_\phi^{(0)} - \xi \right)& \nonumber\\
&\hskip10cm- \frac{1}{z^2 f} \left( \tilde{\rho}^{(0)} \right)^2 \tilde{A}_\phi^{(1)} = 0\ , & 
\end{eqnarray}
and
\begin{eqnarray}
&& \partial_z^2 \tilde{\rho}^{(1)} + \left( \frac{f'}{f} - \frac{2}{z} \right) \partial_z \tilde{\rho}^{(1)} \nonumber\\
&&\hskip1cm+ \frac{1}{f} \left( \partial_{\tilde{r}}^2 \tilde{\rho}^{(1)} + \frac{1}{\tilde{r}} \partial_{\tilde{r}} \tilde{\rho}^{(1)}  - \frac{1}{\tilde{r}^2} \left( \tilde{A}_\phi^{(0)} -\xi \right)^2 \tilde{\rho}^{(1)} - \frac{2}{\tilde{r}^2} \left( \tilde{A}_\phi^{(0)} - \xi \right) \tilde{A}_\phi^{(1)} \tilde{\rho}^{(0)}\right)  \nonumber \\
&&\hskip6cm + \frac{1}{f^2} \tilde{\omega}^2 \tilde{\rho}^{(1)} + \frac{1}{f^2} \tilde{\rho}^{(1)} \left(\tilde{A}_t^{(0)} \right)^2 + \frac{2}{z^2 f} \tilde{\rho}^{(1)} = 0\ .
\end{eqnarray}
What is of particular interest is that we find that the equations for
$\tilde{\rho}^{(1)}$ and $\tilde{A}_\phi^{(1)}$ only depend on each
other, and so we can solve for these two fields independently of
$\tilde{A}_r^{(1)}$ and $\tilde{A}_z^{(1)}$.  For simplicity, we
therefore set these fields to zero.  We now note that under this
assumption, the equation of motion for $\tilde{A}_z^{(1)}$ appears to
be completely independent of $\tilde{A}_r^{(1)}$.  In particular, one
could imagine first solving the equation for $\tilde{A}_z^{(1)}$ and
then substituting the solution into the equation for
$\tilde{A}_r^{(1)}$ and solving for $\tilde{A}_r^{(1)}$ under the
constraint described above.  Unfortunately, we do not know how to
solve the partial differential equation with the constraint, and
therefore we choose to solve \emph{only} for the field
$\tilde{A}_z^{(1)}$.
To proceed, we would like to note that near the event horizon, all the
fields satisfy the same equation:
\begin{equation}
\partial_z^2 X + \frac{f'}{f} \partial_z X + \frac{\tilde{\omega}^2}{f^2} X = 0\ .
\end{equation}
This equation has solutions given by in--going (negative) and
out--going (positive) waves:
\begin{equation}
X  \propto \left( 1 - z \right)^{ \pm i \tilde{\omega} / 3 }\ .
\end{equation}
The correct boundary condition to have at the event horizon is of
in--going waves, so we redefine the field to reflect this:
\begin{equation}
\tilde{A}_z^{(1)} = \left( 1 - z \right)^{- \tilde{\omega}/3} \chi(\tilde{r},z) \ .
\end{equation}
The equation of motion for $Z$ is now given by:
\begin{eqnarray}
&&\partial_z^2 \chi - \frac{\tilde{\omega}^2}{9} \frac{1}{\left(1-z \right)^2} \chi + \frac{f'}{f}  \partial_z \chi + \frac{\tilde{\omega}^2}{f^2} \chi + \frac{i \tilde{\omega}}{3} \left( \frac{2}{1-z} \partial_z \chi + \frac{1}{\left(1-z \right)^2} \chi  + \frac{f'}{f} \frac{1}{1-z} \chi \right)=0 \ ,\nonumber\\
&&\frac{1}{f} \left( \partial_{\tilde{r}}^2 \chi + \frac{1}{ \tilde{r}} \partial_{\tilde{r}} \chi -\left( \tilde{R}^{(0)} \right)^2 \chi  \right)= 0\ . 
\end{eqnarray}
At the event horizon, this gives:
\begin{eqnarray}
&& \partial_{\tilde{r}}^2 \chi + \frac{1}{ \tilde{r}} \partial_{\tilde{r}} \chi   -\left( \tilde{R}^{(0)} \right)^2 \chi  + i \tilde{\omega} \chi + \frac{2}{3} \tilde{\omega}^2 \chi + \left( 2 i \tilde{\omega} - 3 \right) \partial_z \chi = 0 \ , \nonumber\\
&& \partial_z^2 \chi + \left( 2 - \frac{2}{3} i \tilde{\omega} \right) \partial_z \chi + \frac{1}{27} \tilde{\omega} \left( \tilde{\omega} - 6i \right) \chi  = 0\ .
\end{eqnarray}
The first point we need to address when solving these equations is
what the appropriate choice for $\tilde{\omega}$ is.  Since for the
droplet solutions the scalar field asymptotes to zero, we expect that
in order to satisfy the normalizibility conditions for large
$\tilde{r}$, the correct choice for $\tilde{\omega}$ are the
quasi--normal solutions when there is no scalar field present.  We
present some of the values in table \ref{table:complex_omega2}.
\begin{table}[!h] 
\begin{center}
\begin{tabular}{|c|c|} 
\hline $n$ & $\tilde{\omega}$ \\
\hline 0 & 0.66252 -  2.60058 $i$ \\
\hline 1 & 2.20850 -  2.51647 $i$ \\
\hline 2 & 4.03511 -  2.45694 $i$ \\
\hline 3 & 6.06406 -  2.41322 $i$ \\
\hline 4 & 8.26149 -  2.37694 $i$ \\
\hline
\end{tabular}
\caption{Values of $\tilde{\omega}$ that give regularizable solutions for zero scalar field.} \label{table:complex_omega2}
\end{center} 
\end{table}
Using these values, we try to find normalizable solutions for the
droplet scalar field solutions.  What is interesting is that we can
only find solutions for $\Re(\tilde{\omega}) > 4$ (note that this does
not mean that solutions for smaller $\Re(\tilde{\omega})$ do not
exist, just that our extensive numerical search could not find them).
We present an example of this solution in figure
\ref{fig:fluctuations}.  Therefore, these results suggest (note that
this result is not conclusive since we have not solved for the field
$\tilde{A}_r^{(1)}$) that the droplet solutions are stable under
quadratic fluctuations.
The situation for the vortex solutions is very similar, except that
now, the $\tilde{\omega}$ values we use are those of the spatially
independent solutions described earlier.  Once again, we find that
solutions can be easily found above the lower lying modes, although we
find them to be much higher above as illustrated by the example shown
in figure \ref{fig:vortex_fluctuation}.
%
\section{Conclusion}
We have constructed two broad families of localized solution to the
equations of motion for ref.\cite{Hartnoll:2008vx}'s holographic model
of a superconductor, and considered several of their key
properties.

The vortices, with winding number $\xi$, contain $2\pi\xi$ units of
magnetic flux, and are candidates to fill out the superconducting part of
the phase diagram in the presence of an external magnetic field. This
is because a lattice of them in the $(r,\phi)$ plane can trap flux
lines of an applied external $B$--field into filaments as it passes
through the two dimensional sample.  Vortices presumably repel each other, and so such a
lattice will cost energy. Therefore at some critical $B_c(T)$ the
system will seek a lower energy phase, possibly returning to the
normal phase.  We have not constructed such a lattice, and
further study to understand such a configuration 
is a very interesting avenue of research to pursue.  Forming such a lattice is a method by which, at a given $T/T_c<1$, the superconducting
phase can be made to persist in some constant background $B$, even though the
system cannot eject the magnetic field entirely {\it \`a la} the
Meissner effect (there is a nice energetic argument in
ref.\cite{Hartnoll:2008kx} as to why the Meissner effect is not
possible in this two dimensional case). This vortex phase is of course
the same method by which a standard type~II superconductor can persist
beyond the (lower) critical line at which the Meissner effect
disappears, and we expect that it applies here. The study of
multi--vortex solutions needed to establish this is left for further
study.

Crucially, we've established that the droplets do not exist below a
certain critical value of $B$, dropping to zero height on a family of
lines that we were able to compute explicitly. For this and a variety
of other stated reasons, and also considering the fact that they are
of finite size and hence a lattice or gas of them would not give a
connected superconducting path for charge transport, we believe that
they do not represent a superconducting phase. (Hence, we disagree
with the statements made about the phase diagram in
ref.\cite{Hartnoll:2008kx}.  The authors find the critical line, but
state (similarly to our ref.\cite{Albash:2008eh}) that the droplets
exist {\it below} the line, and are superconducting. As they did not
have the full droplet solutions, nor the vortex solutions, their
analyses are not sufficient to make these determinations.). They seem to represent a
non--superconducting phase that is inhomogeneous. Whether or not the
droplets are the favoured solution for arbitrarily large $B$ is an
interesting question. There is the possibility that the system may
prefer to return to the normal phase represented by a dyonic black
hole with zero scalar. Our partial stability analysis showed that the
droplets we studied are stable against fluctuations of the fields, but
there is the possibility that other fluctuation modes may be undamped.
We mention here that we also noticed the curious fact that the droplet
solution at higher magnetic field (that we presented earlier) contains
regions where the local value of the squared scalar mass is below the
Breitenlohner--Freedman bound. 

It would also be of interest to establish whether the critical line
where the vortex phase would disappear coincides with the limiting line
where the droplets' existence begins. We conjecture this to be likely on
the grounds that we have found no other candidate solutions to fill an
intermediate region. While this is the simplest possibility, further
study is needed to establish it firmly.

\section*{Acknowledgments}We would like to thank Arnab Kundu and Rob
Myers for conversations. CVJ thanks the Aspen Center for Physics for a
stimulating working atmosphere while this manuscript was prepared.

\appendix
\section{Normalizations in the Holographic Dictionary} \label{sec:Dictionary}
%
We recall the AdS dictionary (working in Euclidean metric):
\begin{equation}
\left \langle \exp \int \phi_0 \mathcal{O} \right \rangle  = \exp \left(- S_{\mathrm{on-shell}} [ \phi_0 ]  \right) = \mathcal{Z} \ . 
\end{equation}
Taking derivatives on both sides with respect to the boundary source $\phi_0$ gives us our AdS dictionary:
\begin{equation}
\left \langle \mathcal{O} (x_1) \dots \mathcal{O} (x_n) \right \rangle = (\beta {\cal V})^{-n} \lim_{\phi_0 \to 0} \mathcal{Z}^{-1} \frac{\delta }{\delta \phi_0 (x_1)} \dots \frac{\delta }{\delta \phi_0 (x_n)} \mathcal{Z}\ .
\end{equation}
We define the the free energy density of the dual theory to be given:
\begin{equation}
\mathcal{F} = \frac{1}{\beta {\cal V}} S_{\mathrm{on-shell}} \ .
\end{equation}
where $\beta$ is the inverse temperature and ${\cal V}$ is the ``spatial volume'' of the dual theory.  If we use the notation that we are working in AdS$_{d+1}$, then ${\cal V}$ has mass dimension $d-1$.  For us, $d=3$.  In the Euclidean language, our action is given by:
\begin{equation}\label{eqt:action}
S_{\mathrm{bulk}} = \frac{1}{2 \kappa_4^2} \int d^4 x \sqrt{-G} \left\{ -R - \frac{6}{L^2} + L^2 \left( \frac{1}{4} F^2 + \left| \partial \Psi - i g A \Psi \right|^2 + {\cal V} \left(\left|\Psi \right| \right)\right) \right\}\ .
\end{equation}
where we emphasize that although the metric is now purely positive, we are still using $A_t$ and now a Wick rotated version of it.  In the dual theory, the charge density is given by:
\begin{equation}
\rho(x) = - \frac{\delta \mathcal{F}}{\delta \mu(x)}\ .
\end{equation}
where $\mu$ is the chemical potential and $x$ represents the space--time coordinates in the dual field theory.  Using the AdS/CFT dictionary, we can write:
\begin{equation}
\frac{\delta \mathcal{F}}{\delta \mu(x)} = 
\frac{1}{\beta {\cal V}} \frac{\delta S_{\mathrm{on-shell}}}{\delta A_t (x,0)} \ .
\end{equation}
Using the action given in equation \reef{eqt:action}, we find:
\begin{equation} \label{eqt:charge_density}
\frac{g}{\beta {\cal V}} \frac{\delta S_{\mathrm{on-shell}}}{\delta A_t (x,0)} =   \frac{L^2}{2 \kappa_4^2} \frac{1}{g} \alpha \partial_z A_t (x, 0) =  \frac{L^2}{2 \kappa_4^2} \frac{1}{g} \alpha^2 \partial_z \tilde{A}_t (x, 0)  \ ,
\end{equation}
where we have used that:
\begin{equation}
\frac{ \delta A_t(x')}{\delta A_t (x)} = \beta {\cal V} \delta^{(d+1)} \left(x' - x  \right) \ .
\end{equation}
and we have dropped the contribution coming from the event horizon.  Therefore, the end result is given by:
\begin{equation}
\rho(x) =  -   \frac{L^2}{2 \kappa_4^2} \frac{1}{g} \alpha^2 \partial_z \tilde{A}_t (x, 0) \ .
\end{equation}
A similar procedure allows us to calculate the vev of the azimuthal current as well:
\begin{equation}
J^\phi (x)= - g \beta {\cal V}  \frac{ \delta S_{\mathrm{on-shell}}}{\delta A_\phi (x)} = \frac{L^2}{2 \kappa_4^2} \frac{1}{g} \frac{\alpha^3}{\tilde{r}^2} \partial_z  \tilde{A}_\phi\ .
\end{equation}
Note that here we are using the vector field that has been rescaled by
$g$, hence why the factor of $g$ appears at the beginning of our
definition.  Next, we can calculate the form of the vevs of the
$\Delta = 1$ and $\Delta = 2$ operators.  To proceed, we take the
$\Delta = 1$ operator to be the source of the $\Delta = 2 $ operator
\cite{Klebanov:1999tb}.  Therefore, we write:
\begin{equation}
\langle \mathcal{O}_2 (x) \rangle = -  \frac{ \delta \mathcal{F}}{ \delta \langle \mathcal{O}_1 (x) \rangle } \propto  - \frac{ \delta \mathcal{F}}{ \delta \rho_1 (x)} \ ,
\end{equation}
where we are using the notation that:
\begin{equation}
\rho(x, z \to 0) = z \rho_1(x) + z^2 \rho_2 (x) \ .
\end{equation}
To proceed, we calculate the variation of the bulk action (keeping
only divergent and finite terms):
\begin{equation}
\delta S_{\mathrm{bulk}} = - \lim_{z \to 0} \frac{L^2}{2 \kappa_4^2} \frac{1}{g^2} \int d^3 x \  L^2 \alpha^3\left( \frac{1}{z} \rho_1 (x) \delta \rho_1(x) + 2  \rho_2 (x) \delta \rho_1 (x)  + \rho_1(x) \delta \rho_2(x) \right) \ .
\end{equation}
The first term in parentheses is divergent, but it is removed by an
appropriate counterterm:
\begin{equation}
S_{\mathrm{CT}} = -\frac{L^2}{2 \kappa_4^2}  \frac{1}{g^2} \lim_{z \to 0}  \frac{\sqrt{-\gamma}}{L} \int d^3 x \ \frac{1}{2} \rho (x, z)^2  \ ,
\end{equation}
which leaves us with:
\begin{equation}
\delta S_{\mathrm{bulk}}  + \delta S_{\mathrm{CT}}  = - \frac{L^2}{2 \kappa_4^2} \frac{1}{g^2} \int d^3 x \  L^2 \alpha^3  \rho_2 (x) \delta \rho_1 (x) \ .
\end{equation}
Therefore, we find:
\begin{equation}
-\frac{1}{\alpha L}  \frac{ \delta \mathcal{F}}{ \delta \rho_1 (x)} = \frac{L^2}{2 \kappa_4^2} \frac{1}{g^2} L \alpha^2 \rho_2 (x)  =  \frac{L^2}{2 \kappa_4^2} \frac{1}{g^2} \alpha^2 \tilde{\rho}_2 (x) \ .
\end{equation}
To calculate the vev of the $\Delta = 1$ operator, we need to perform
a Legendre transform on $\mathcal{F}$ \cite{Klebanov:1999tb}:
\begin{equation}
\mathcal{G} =  -\mathcal{F} - \frac{1}{\beta {\cal V}} \frac{L^2}{2 \kappa_4^2} \frac{L^2 \alpha^3}{g^2} \int d^3 x \ \rho_1(x)  \rho_2 (x) \ ,
\end{equation}
and we now have:
\begin{equation}
-\frac{1}{L \alpha^2} \frac{\delta \mathcal{G}}{ \delta \rho_2 (x) } =  \frac{L^2}{2 \kappa_4^2} \frac{1}{g^2} L \alpha \rho_1 (x) =  \frac{L^2}{2 \kappa_4^2} \frac{1}{g^2} \alpha \tilde{\rho}_1 (x) \ .
\end{equation}
Therefore, in order to satisfy the conditions:
\begin{equation}
\langle \mathcal{O}_2(x) \rangle = -\frac{ \delta \mathcal{F}}{\delta \langle \mathcal{O}_1(x) \rangle} \ , \quad \langle \mathcal{O}_1 (x)\rangle = -\frac{ \delta \mathcal{G}}{\delta \langle \mathcal{O}_2 (x)\rangle} \ ,
\end{equation}
we choose:
\begin{eqnarray}
\langle \mathcal{O}_1 (x) \rangle &=&  \frac{L}{\sqrt{2} g \kappa_4}  L \alpha \rho_1(x) =    \frac{L}{\sqrt{2} g \kappa_4} \alpha \tilde{\rho}_1 (x) \ , \\
\langle \mathcal{O}_2 (x) \rangle &=&   \frac{L}{\sqrt{2} g \kappa_4}  L \alpha^2 \rho_2 (x)=    \frac{L}{\sqrt{2} g \kappa_4} \alpha^2 \tilde{\rho}_2 (x) \ .
\end{eqnarray}
%
\section{Flux Quantization} \label{sec:flux_quantization}
%
Let us review the flux quantization condition.  Deep in the
superconductor, away from the vortex, our results indicate that we
have $\tilde{A}_\phi = \xi$.  This can be written as:
\begin{equation} \label{eqt:sc_cond}
\frac{1}{r} A_\phi=\frac{1}{r} \partial_\phi \theta\ .
\end{equation}
It is useful at this point to realize that $A_\phi / r $ transforms
exactly as the vector $\vec{A}$.  Therefore, it is convenient to write
equation \reef{eqt:sc_cond} as:
\begin{equation}
\vec{\nabla}{\theta} =  \vec{A}\ .
\end{equation}
Consider drawing a circle around the vortex, deep in the
superconductor.  We can choose to integrate equation
\reef{eqt:sc_cond} along the circle:
\begin{equation}
\oint \vec{\nabla} \theta  \cdot d \vec{\ell} = \oint \vec{A}  \cdot d \vec{\ell}\ .
\end{equation}
where $A$ is the corresponding 1--form.  The left hand side gives $2
\pi \xi$, and using Stokes' theorem on the RHS, we have:
\begin{equation}
2 \pi \xi = \int \left(\vec{\nabla} \times \vec{A}\right)_z d a =  \int d a \left(\frac{1}{r} \partial_r A_\phi\right)\ .
\end{equation}
Note that the quantity in brackets on the right is exactly the
magnetic field, so we find that the flux of a single vortex is
quantized to $2 \pi \xi$.
%
\providecommand{\href}[2]{#2}\begingroup\raggedright\endgroup

\end{document}